\documentclass[twocolumn,showpacs,amsmath,nofootinbib,pra,aps,amssymb,longbibliography]{revtex4-1}

\usepackage[T1]{fontenc}
\usepackage[utf8]{inputenc}
\usepackage{times}
\usepackage{color} 
\usepackage{array}
\usepackage{amssymb,amsmath}
\usepackage{amsbsy}
\usepackage[pdftex]{graphicx}
\usepackage{bm} 
\usepackage{float} 
 
\usepackage[unicode,breaklinks]{hyperref}
\hypersetup{
    unicode=true,
    plainpages=false, 
    colorlinks=true,
    linkcolor=blue,
    citecolor=blue,
    filecolor=black,
    urlcolor=blue
}
\usepackage{url} 
\usepackage{verbatim}

\newcommand{\gdd}{g_{dd}}

\newcommand{\br}{\bm{r}}
\newcommand{\brho}{{\boldsymbol{\rho}}}
\newcommand{\bx}{\bm{x}}

\newcommand{\gQF}{g_{\mathrm{QF}}}

\newcommand{\Ut}{\tilde{U}}
\newcommand{\Lz}{\mathcal{L}_z}

\newcommand{\LD}{\mathcal{L}_\mathrm{3D}}

\DeclareMathOperator{\Ei}{Ei}

\begin{document}

\title{Variational theory for the ground state and collective excitations of an elongated dipolar condensate}
	\author{P.~Blair~Blakie, D.~Baillie, and Sukla Pal}
	\affiliation{%
	 Dodd-Walls Centre for Photonic and Quantum Technologies, New Zealand and Department of Physics, University of Otago, Dunedin 9016, New Zealand}	
\date{\today}
\begin{abstract}  
We develop a variational theory for a dipolar condensate in an elongated (cigar shaped) confinement potential. Our formulation provides an effective one-dimensional extended meanfield theory for the ground state and its collective excitations.  We apply our theory to investigate the properties of rotons in the system comparing the variational treatment to a full numerical solution. We consider the effect of quantum fluctuations on the scattering length at which the roton excitation softens to zero energy.
\end{abstract} 


\maketitle

\section{Introduction}
\begin{figure}[htbp] 
   \centering
   \includegraphics[width=3.4in]{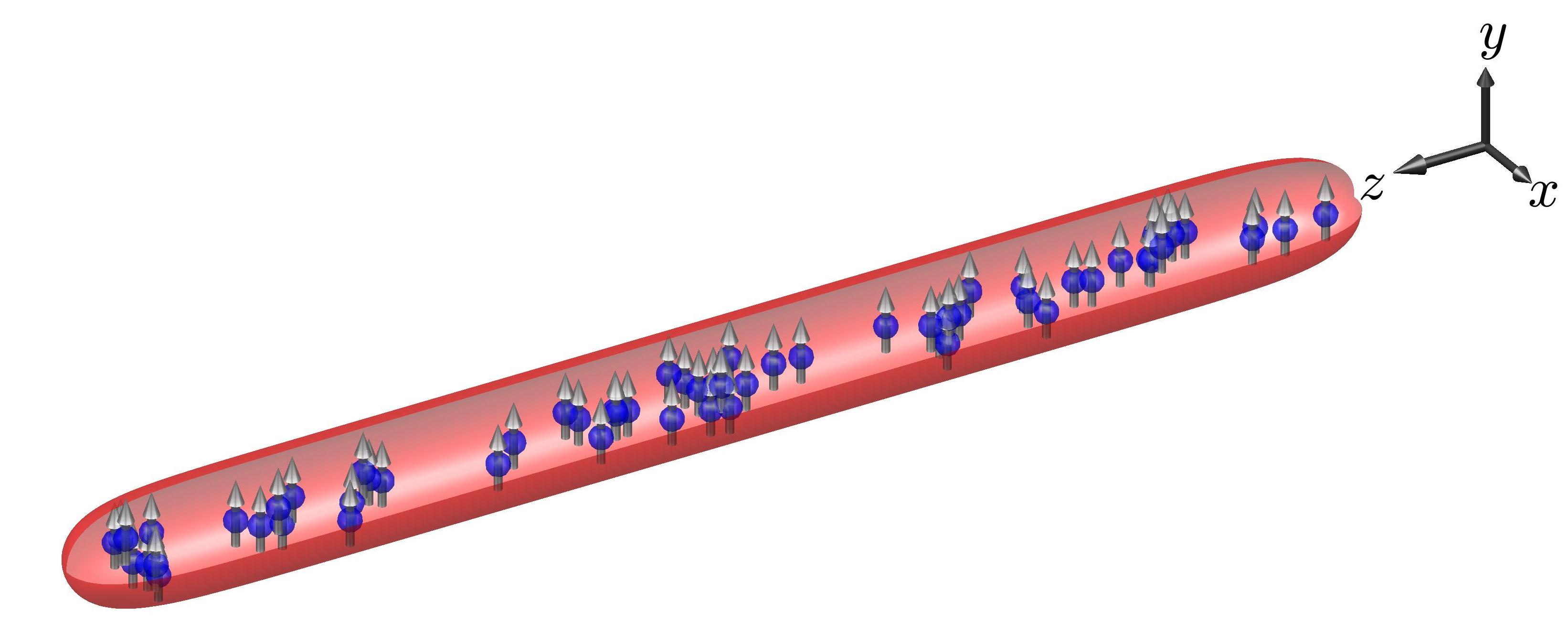}  
   \caption{ Schematic of the system geometry we consider in this paper: A condensate of atoms with magnetic dipoles aligned along the $y$-axis by an external magnetic field, and with tighter confinement in the $xy$-plane relative to the $z$-axis.  
   }
   \label{fig:schematic}
\end{figure}

 Bose-Einstein condensates (BECs)  of highly magnetic atoms, such as chromium, erbium, and dysprosium \cite{Griesmaier2005a,*Pasquiou2011a,Mingwu2011a,*Lu2012a,Aikawa2012a}, realize a dilute quantum system with long-ranged and anisotropic dipole-dipole interactions (DDIs). Recently experiments with such dipolar condensates have observed a roton excitation \cite{Chomaz2018a,Petter2018a}, i.e.~a local minimum in the excitation dispersion relation of the condensate at a non-zero wave vector. The roton excitation, originally introduced in the study of super fluid Helium \cite{Landau1941a}, has been of significant theoretical interest in dipolar condensates where it emerges from an interplay between the DDIs and confinement (e.g.~see \cite{Santos2003a,Ronen2007a,Blakie2012a,Corson2013a,Corson2013b,JonaLasinio2013b,Bisset2013a,Baillie2015b,Roccuzzo2019a}). Much of the theoretical attention has focused on the case of a system confined to a planar (pancake) geometry with the dipoles aligned along the tightly confined direction. Experiments instead have realized roton excitations in an elongated (cigar) geometry with the dipoles oriented along one of the tightly confined directions (e.g.~see Fig.~\ref{fig:schematic}). While the planar case can have cylindrical symmetry, the elongated system does not and in general requires a full three-dimensional (3D) calculation. Such calculations for the system ground states and its collective excitations demand significant computational resources, mostly arising from the large and dense numerical grids required to carefully resolve the singular DDI potential. 
 
 The theoretical description of  a dipolar condensate is usually provided by meanfield theory. However, recent developments in the field have revealed that quantum fluctuations can play an important role in the regime of strong DDIs, for example leading to the formation of stable quantum droplets and supersolids (e.g.~see \cite{Kadau2016a,Ferrier-Barbut2016a,Bisset2016a,Bisset2016a,Chomaz2016a,Schmitt2016a,Bottcher2019b,Bottcher2019a,Tanzi2019a,Chomaz2016a}). Extended meanfield theory includes the leading order effects of quantum fluctuations within a local density approximation \cite{Ferrier-Barbut2016a,Wachtler2016a,Bisset2016a}. In this formalism the stationary states of the system are described by the extended Gross-Pitaevskii equation (eGPE), with the collective excitations described by the associated  Bogoliubov-de Gennes (BdG) equations  \cite{Baillie2017a,Lee2018a}.

 In this paper we develop an approximation that allows us to reduce the 3D extended meanfield theory to a tractable one-dimensional (1D) form. Our approach is to use a Gaussian ansatz to describe the tightly confined transverse directions, and by integrating this out we obtain an effective 1D form of the theory, albeit with some variational parameters from the Gaussian. Such an approach is a rather obvious path to take and has been used for non-dipolar condensates (e.g.~see \cite{Salasnich2002a}). However, for the dipolar case the Gaussian cannot be integrated out against the DDI potential to yield an analytic result for the interaction term, except for the special case where the Gaussian is isotropic. An important result of this paper is that we introduce a useful approximate analytic result for general Gaussian.  Based on this we develop a simple 1D  effective theory for the stationary states and the collective excitations of the system.  We emphasize that this description is for a 3D dipolar condensate, and is not applicable to the true 1D regime where interactions remain weak compared to the confinement  and the quantum fluctuations take a different form \cite{Edler2017a}.
 Indeed, in  the regime of interest (e.g.~where rotons occur) the interaction energy scale is typically larger than the transverse confinement energy and the transverse degrees of freedom must be treated variationally. Furthermore, in this regime magnetostriction effects can be large, causing the Gaussian to distort appreciably from the geometry imposed by the confining potential. This occurs because the DDIs are anisotropic and the energy of the system is reduced by having more particles in a relative head-to-tail orientation.

 We compare our results from the variational 1D theory to full numerical solutions of the 3D problem for both ground states and the excitation spectrum.  
  We use our theory to predict the value of the s-wave scattering length (readily adjusted in experiments using Feshbach resonances) where the roton mode goes soft (i.e.~to zero energy), and the associated value of the wave vector of the soft mode. By comparing to results that exclude the quantum fluctuation term we can assess the effect of quantum fluctuations on the roton properties.  
 
The outline of the paper is as follows. In Sec.~\ref{Sec:Formalism} we introduce extended meanfield theory and our approach for simplifying it to an effective variational 1D theory.   The main results are presented in Sec.~\ref{sec:Results}, before we conclude in Sec.~\ref{Sec:Conclusions}.

\section{Formalism}\label{Sec:Formalism}
\subsection{Extended meanfield theory for dipolar condensates}
\subsubsection{3D Extended Gross-Pitaevskii equation (eGPE)} 
The system of interest is a dilute Bose gas of atoms with a magnetic moment $\mu_m$ polarized along the $y$-axis. The extended meanfield theory for this system identifies stationary states of the matterwave field $\Psi_0$ as solutions of the eGPE $\mu\Psi_0=\LD\Psi_0$, where 
\begin{equation}
\LD\Psi_0=\left[ -\frac{\hbar^2 \nabla^2}{2m} + V(\bx) + \Phi(\bx) + \gamma_{\mathrm{QF}} |\Psi_0|^3\right]\Psi_0.\label{e:GPE}
\end{equation}
Here $\mu$ is the chemical potential and
\begin{align}
  \Phi(\bx) = \int \text{d}\bx'\,U(\bx-\bx')|\Psi_0(\bx')|^2, \label{Phi3D}
  \end{align}
is the interaction term, with interaction potential
\begin{align}
U(\br) = g_s\delta(\br) + \frac{3\gdd}{4\pi r^3}\left(1-3\frac{y^2}{r^2}\right),\label{U3Dr}
\end{align}
 where
 the contact interaction coupling constant is $g_s = 4\pi\hbar^2a_s/m$,   $a_s$ is the $s$-wave scattering length, the DDI coupling constant is
$\gdd = 4\pi\hbar^2a_{dd}/m$, and  $a_{dd}\equiv m\mu_0\mu_m^2/12\pi\hbar^2$ is the dipole length.
This theory includes the leading order quantum fluctuations in the local density approximation, where the coefficient of this term is \cite{Lima2011a,Wachtler2016a,Bisset2016a}
\begin{align}
\gamma_{\mathrm{QF}} = \frac{32}{3}g_s\sqrt{\frac{a_s^3}{\pi}}(1+\tfrac32 \epsilon_{dd}^2),
\end{align}
with  $\epsilon_{dd} =a_{dd}/a_s$. The atoms are taken to be confined by an external potential $V(\bm{x})=\frac{1}{2}m\sum_{\nu=x,y,z}\omega_\nu^2 \nu^2$.
Here we focus our attention on the regime used in experiments to observe roton excitations: a cigar shaped trap with $\omega_x,\omega_y \gg\omega_z$ (including the pure tube case with $\omega_z=0$). The energy  functional associated with the 3D eGPE is
\begin{equation}
E=\int \text{d}\bx \,\Psi_0^*\left[ -\frac{\hbar^2 \nabla^2}{2m} + V(\bx) + \tfrac12\Phi(\bx) + \tfrac25\gamma_{\mathrm{QF}} |\Psi_0|^3\right]\Psi_0.\label{e:E3d}
\end{equation}

\subsubsection{Bogoliubov-de Gennes theory of excitations}
The collective excitations of this system are Bogoliubov quasiparticles, which can be obtained by linearizing the time-dependent GPE $\text{i}\hbar\dot{\Psi}=\LD\Psi$ about a stationary state as 
\begin{align}
\Psi\!=\!\text{e}^{-\text{i}\mu t/\hbar}\left[\Psi_0\!+\!\sum_\nu\left( \lambda_\nu U_{\nu}\text{e}^{-\text{i}\epsilon_\nu t/\hbar}-\lambda_\nu^*V_{\nu}^*\text{e}^{\text{i}\epsilon_\nu^* t/\hbar}\right)\right]\!\!,\! 
\end{align}
where $\lambda_\nu$ is a small complex amplitude.
The quasiparticle modes  ${U_\nu,V_\nu}$ and energies $\epsilon_\nu$ satisfy the  BdG  equations \cite{Baillie2017a}
\begin{align}
    \!\!\!\begin{pmatrix}  \LD - \mu + X & -X \\
      X &\! -(\LD - \mu + X)\end{pmatrix}\!\begin{pmatrix} U_\nu \\ V_\nu\end{pmatrix}
     &= \epsilon_\nu \!\begin{pmatrix} U_\nu \\ V_\nu\end{pmatrix},\label{BdG3D}\end{align}
where $X$ is the exchange operator given by
\begin{align}
  Xf &\equiv\Psi_0\!\int \!\text{d}\bx'U(\bx\!-\!\bx') f(\bx')\Psi_0^*(\bx')
  \!+\! \tfrac32 \gamma_{\mathrm{QF}} |\Psi_0|^3f.
\end{align}

\subsection{Reduction to an effective 1D eGPE} 
\subsubsection{General approach}
We approximate the 3D solutions in the elongated trap to be of the separable form $\Psi_0(\bm{x})=\psi_0(z)\chi(\brho)$, where  $\chi$ is the  transverse mode function with $\brho=(x,y)$ being the radial coordinate vector and $\int \text{d}\brho\,|\chi|^2=1$. Integrating out the transverse directions we obtain the 1D eGPE operator for the axial wavefunction $\psi_0$:
\begin{align}
\Lz&=\int \text{d}\bm{\rho}\,\chi^*\LD\chi,\\
    &= \mathcal{E}_\perp-\frac{\hbar^2}{2m}\frac{\text{d}^2}{\text{d}z^2} +\tfrac{1}{2}m\omega_z^2z^2 +\Phi_z(z)+\gamma_{\mathrm{QF}}\gamma_\perp|\psi_0|^3,
\end{align}
where $\gamma_\perp\equiv\int \text{d}\bm{\rho}|\chi|^5$, and
\begin{align}
\mathcal{E}_\perp&=\int \text{d}\bm{\rho}\,\chi^*\left[-\frac{\hbar^2\nabla_{\bm{\rho}}^2}{2m}+\tfrac{1}{2}m(\omega_x^2x^2+\omega_y^2y^2)\right]\chi.
\end{align}
The  effective interaction term is
\begin{align}
\Phi_z(z)&=\mathcal{F}_z^{-1}\left\{\Ut_z(k_z )\mathcal{F}_z\{|\psi_0|^2\}\right\}\label{Phiz},
\end{align}
where we have introduced the effective 1D $k$-space interaction kernel
\begin{align}
    \Ut_z(k_z)&=\int \frac{\text{d}\bm{k}_\rho}{(2\pi)^2}\Ut(\bm{k})\left|\mathcal{F}_{\bm{\rho}}\{|\chi|^2\}\right|^2.\label{Uz}
\end{align}
In the above results  $\Ut(\bm{k})$ is the Fourier transform of Eq.~(\ref{U3Dr}), given by
\begin{align} 
\Ut(\bm{k})=g_{s}+g_{dd}\left(3\frac{k_y^2}{k^2}-1\right),\label{U3Dk}
\end{align}
$\mathcal{F}_z\{f\}$ denotes the 1D Fourier transform $f(z)\to \tilde{f}(k_z)$  and $\mathcal{F}_{\bm{\rho}}\{f\}$ denotes the 2D Fourier transform  $f(\bm{\rho})\to\tilde{f}(\bm{k}_\rho)$.  
The associated energy functional takes the form
\begin{align}
E=&\int \text{d}z\,\psi_0^*\left[\mathcal{E}_\perp  -\frac{\hbar^2 }{2m}\frac{\text{d}^2}{\text{d}z^2} + \tfrac{1}{2}m\omega_z^2z^2\right]\psi_0\nonumber \\
&+\int \text{d}z\,\psi_0^* \left[\tfrac12\Phi_z(z) + \tfrac25\gamma_{\mathrm{QF}}\gamma_\perp |\psi_0|^3\right]\psi_0.\label{e:E1d}
\end{align}

\subsubsection{Anisotropic Gaussian approximation}\label{Sec:AnisoGaussian}
Here we introduce a convenient analytic form for $\chi$. Our choice is the Gaussian 
\begin{align}
\chi_\sigma(\brho)=\frac{\text{e}^{-(\eta x^2+y^2/\eta)/2l^2}}{\sqrt{\pi}l},\label{chi_eta}
\end{align}
 of mean width $l=\sqrt{l_xl_y}$, and anisotropy  $\eta=l_y/l_x$, where  $l_x$ ($l_y$) is the 1/e half width of $|\chi_\sigma|^2$ along the $x$-axis ($y$-axis).  We use $\sigma$ to collectively denote the variational parameters $\sigma=\{l,\eta\}$, which are determined by minimising the system energy. 
 
Using $\chi_\sigma$  we can analytically evaluate  key terms in the 1D theory. First we denote $\mathcal{E}_\perp$ evaluated with $\chi_\sigma$ as  
 \begin{align}
 \mathcal{E}_\sigma(l,\eta)= \frac{\hbar^2}{4ml^2}\left(\eta+\frac{1}{\eta}\right)+\frac{ml^2}{4}\left(\frac{\omega_x^2}{\eta}+\omega_y^2\eta\right). \end{align}
 Similarly,  $\gamma_{\perp}\to\gamma_\sigma=\frac{2}{5\pi^{3/2}l^3}$. We are unaware of a general analytic result for $\tilde{U}_z(k_z)$  evaluated using $\chi_\sigma$, which we denote as $\Ut_\sigma(k_z)$.  However, we have obtained the useful approximate  result 
\begin{align}
\Ut_\sigma(k_z)&=\frac{g_s}{2\pi l^2}\! +\!\frac{g_{dd}}{2\pi l^2}\!\left\{\frac{3  [  Q_\sigma^2\text{e}^{ Q_\sigma^2}\Ei(- Q_\sigma^2) +1 ]}{1+\eta}-1\right\},\label{Ueta}
\end{align}
with  $\Ei$ being the exponential integral\footnote{ {This can also be written in terms of the incomplete Gamma function $\Gamma$ using $\Ei(-x)=-\Gamma(0,x)$ for $x>0$ (cf.~Ref.\cite{Sinha2007a}).}}, and $Q_\sigma\equiv \tfrac{1}{\sqrt{2}}k_z\eta^{1/4}l$.

\subsubsection{Justification for Eq.~(\ref{Ueta})}\label{Sec:justification}

\begin{figure}[htbp] 
   \centering
   \includegraphics[width=3.2in]{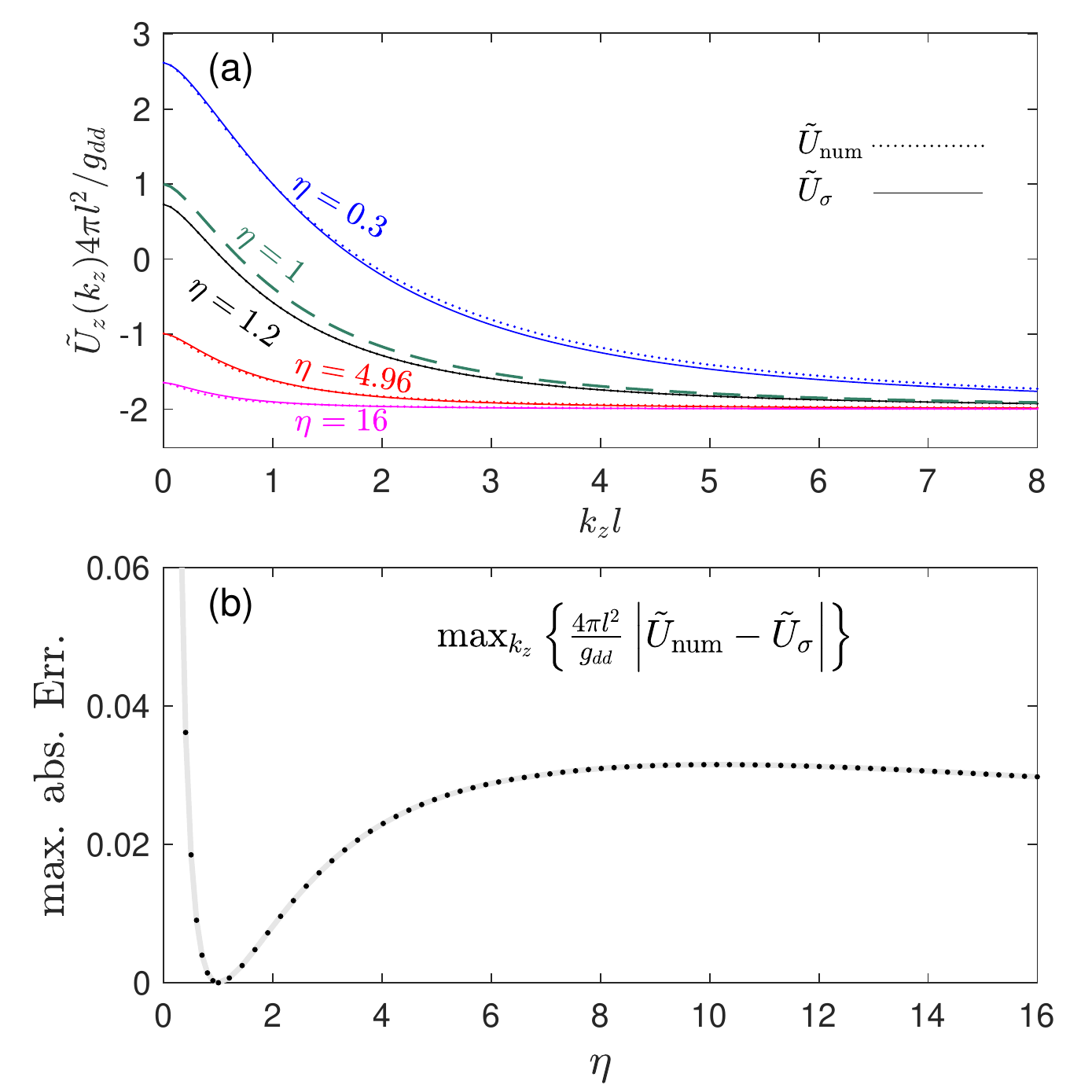}  
   \caption{(a) Comparison of the (dotted lines) analytic result (\ref{Ueta}) to (solid lines) numerically calculated $\tilde{U}_{\mathrm{num}}$   [obtained by numerically evaluating Eq.~(\ref{Uz}) using $\chi\to\chi_\sigma$]  for the effective $k$-space kernel. Results shown for several values of $\eta$ and for $g_s=0$. The exact result for $\eta=1$ is also shown (dashed line). (b) The maximum absolute error of the approximation $\tilde{U}_\sigma$  compared to the $\tilde{U}_{\mathrm{num}}$ over the $k_z$-range shown in (a), (in units of $g_{dd}/4\pi l^2$). }
   \label{fig:GaussianEta}
\end{figure}

For the particular case of an isotropic Gaussian (i.e.~$\eta=1$) Eq.~(\ref{Ueta}) is exact (see \cite{Deuretzbacher2010a,Deuretzbacher2013a,Giovanazzi2004a,Sinha2007a}).
While a general analytic result for $\eta\ne1$ is unavailable, we can calculate the limiting behavior 
\begin{align}
\Ut_{\sigma}(k_z)=\frac{g_s}{2\pi l^2}+
\left\{\begin{array}{cll}
      \frac{g_{dd}}{2\pi l^2}\frac{2-\eta}{1+\eta}, & &  k_z\to0\\
      &\\
        -\frac{g_{dd}}{2\pi l^2}, &  & k_z\to \infty
        \end{array}\right.
\end{align}
We have arrived at result (\ref{Ueta}) by inspection and numerical experiment: it  satisfies the required limiting behavior and reduces to the exact isotropic result at $\eta=1$.

 In Fig.~\ref{fig:GaussianEta} we compare the accuracy of Eq.~(\ref{Ueta}) to a full calculation of the kernel obtained by numerically evaluating (\ref{Uz}) with  $\chi\to\chi_\sigma$. To ensure the numerical calculation is accurate it is performed using a large and dense two-dimensional transverse grid of points and using a cutoff $k$-space DDI potential to avoid finite size boundary effects (e.g.~see \cite{Ronen2006a}). 
 The results in Fig.~\ref{fig:GaussianEta} show that while our approximate analytic result (\ref{Ueta}) is not identical to the numerical result, it is generally in very good agreement over a wide range of $\eta$ values (i.e.~$0.2\lesssim\eta\lesssim20$).  {Note that ground states with $\eta<1$ can occur due to the confinement (i.e.~when $\omega_y>\omega_x$), and are also favoured by the interactions when $g_{dd}<0$, which can be arranged by rotationally tuning the dipoles \cite{Giovanazzi2002b,Tang2018a}.} We expect that for the regimes of interest the error associated with making the Gaussian approximation is much more significant than any additional error introduced by using Eq.~(\ref{Ueta}) to describe its interactions.

\subsubsection{Variational theory}\label{Sec:1DeGP}
Here we summarise the results developed in Sec.~\ref{Sec:AnisoGaussian} and succinctly present the variational 1D eGPE theory that forms the main formalism result of this paper.  

The axial orbital $\psi_0$ satisfies the 1D eGPE:
\begin{align}
\mu\psi_0=\mathcal{L}_\sigma\psi_0,\label{eGPE1Dvar}
\end{align}
where
\begin{align}
\mathcal{L}_\sigma= \mathcal{E}_\sigma-\frac{\hbar^2}{2m}\frac{\text{d}^2}{\text{d}z^2} +\tfrac{1}{2}m\omega_z^2z^2 +\Phi_\sigma(z)+g_{\mathrm{QF}}|\psi_0|^3,
\end{align}
with $g_{\mathrm{QF}}=\gamma_{\mathrm{QF}}\gamma_\sigma$, and $\Phi_\sigma$ is evaluated according to  Eq.~(\ref{Phiz}) but using $\tilde{U}_\sigma$ in place of $\tilde{U}_z$.

Since $\mathcal{L}_\sigma$ depends on $\chi_\sigma$ we also need a procedure to obtain the parameters $\{l,\eta\}$. To do this we consider the total system energy per particle: 
\begin{align} 
\mathcal{E}[\psi_0;l,\eta]=\mathcal{E}_\sigma(l,\eta)+\mathcal{E}_z[\psi_0;l,\eta],\label{eGPE_Efun}
\end{align}
where
\begin{align}
\mathcal{E}_z[\psi_0;l,\eta]=&\frac{2}{5N}\!\int \text{d}z\, g_{\mathrm{QF}}|\psi_0|^5\,\label{Ez}\\
 +\frac{1}{N}\!\int \text{d}z\,&\psi_0^*\left(\!-\frac{\hbar^2}{2m}\frac{\text{d}^2}{\text{d}z^2}+\tfrac{1}{2}m\omega_z^2z^2+\tfrac{1}{2}\Phi_\sigma(z)\right)\psi_0\nonumber
\end{align}
is the energy functional for the $\psi_0$ orbital, and $N=\int \text{d}z\,|\psi_0|^2$.

For  $\omega_z=0$ the axial wavefunction can be uniform $\psi_0\to\sqrt{n}$, where  $n$ is the linear density. In this regime the energy per particle (\ref{eGPE_Efun}) reduces to  
\begin{align}
\mathcal{E}_u(l,\!\eta) &=\mathcal{E}_\sigma(l,\!\eta)+ \tfrac{1}{2}n \Ut_\sigma(0)+\tfrac{2}{5}g_{\mathrm{QF}}n^{{3/2}},\label{Eunif}
\end{align}  
i.e.~the ground state is determined by minimising a simple nonlinear function.

\subsubsection{Quasi-1D theory}\label{Sec:q1D}

Predictions for $\psi_0$ can be made within the quasi-1D approximation using the procedure outlined for the variational theory, but with $l$ and $\eta$ held fixed to the values for the harmonic oscillator ground state of the transverse confinement, i.e.~for $l^2=  {\hbar}/{m\sqrt{\omega_x\omega_y}}$ and $\eta=\sqrt{\omega_x/\omega_y}$, with $\mathcal{E}_\sigma=\hbar(\omega_x+\omega_y)/2$.  We denote the harmonic oscillator state as $\chi_{\mathrm{ho}}$ and the associated $k$-space kernel as $\Ut_{\mathrm{ho}}$. 
This quasi-1D approximation will only be accurate when the interaction terms in $\mathcal{E}_z$ remain small compared to $\mathcal{E}_\sigma$.  

\subsubsection{Effective 1D form of the excitations}\label{Sec:Excitations} 
Making the same shape approximation  for the transverse form of the excitations \cite{Baillie2015b} we set $U_\nu(\bm{x})=u_\nu(z)\chi_\sigma(\bm{\rho})$ and $V_\nu(\bm{x})=v_\nu(z)\chi_\sigma(\bm{\rho})$ and integrating out $\chi_\sigma(\bm{\rho})$ the BdG equations (\ref{BdG3D}) reduce to
\begin{align}
 \! \begin{pmatrix} \mathcal{L}_\sigma-\mu+X_\sigma & -X_\sigma \\ X_\sigma & -(\mathcal{L}_\sigma-\mu+X_\sigma)
  \end{pmatrix}
  \begin{pmatrix}
      u_\nu \\ v_\nu
  \end{pmatrix}
  =
  \epsilon_\nu
  \begin{pmatrix}
      u_\nu \\ v_\nu
  \end{pmatrix},\label{BdG}
\end{align}
where $X_\sigma f =  \psi_0 \mathcal{F}_z^{-1}\left\{ \tilde{U}_\sigma(k_z)\mathcal{F}_z\{ \psi_0f\}\right\} +\frac32 \gQF \psi_0^3f $.

In general the BdG equations need to be discretised and solved numerically, however for the case of a uniform ground state an analytic solution can be obtained. Here the excitations are plane waves of momentum $\hbar k_z$, i.e.~$u_\nu(z)\to \mathrm{u}_{k_z}\text{e}^{\text{i}k_zz}$, $v_\nu(z)\to \mathrm{v}_{k_z}\text{e}^{\text{i}k_zz}$, $\epsilon_\nu \to\epsilon(k_z)$, with excitation energy
\begin{align}
\epsilon(k_z)= \sqrt{\epsilon_0(k_z)\left[\epsilon_0(k_z)+2n\Ut_\sigma(k_z)+3\gQF n^{3/2}\right]} ,\label{BdGUniform}
\end{align} 
where $\epsilon_0(k_z)=\hbar^2k_z^2/2m$. 

\section{Results}\label{sec:Results}

 \subsection{Numerical Methods} 
 In this subsection we briefly outline the various numerical methods used to solve for the results we present later.
 
\subsubsection{Uniform cases}
For cases without axial trapping ($\omega_z=0$) we restrict our attention to the regime where the ground state is uniform and specified by the linear density $n$. 

The variational 1D eGPE theory reduces to minimising the nonlinear function (\ref{Eunif}) for $l$ and $\eta$. The BdG excitation energies are then directly given by evaluating Eq.~(\ref{BdGUniform}).

The 3D eGPE reduces to the determining the transverse mode $\chi(\bm{\rho})$. We do this by discretizing $\chi(\bm{\rho})$ on a two dimensional numerical grid and apply discrete Fourier transformations to apply the kinetic energy operator with spectral accuracy, and to evaluate the interaction term $\Phi$. For high accuracy the  3D $k$-space kernel is cutoff in the transverse direction to the range of the numerical grid (e.g.~see \cite{Ronen2006a,Lu2010a}). The eGPE is solved using a gradient flow technique \cite{Bao2010a}.  The excitations for this case are of the form of plane waves along $z$,  reducing the BdG equations to a 2D form that can be solved using large-scale eigensolvers (i.e.~the implicitly restarted Arnoldi method) .

\subsubsection{Fully trapped cases}
For $\omega_z\ne0$ the variational theory (including the quasi-1D theory) involves solving for $\psi_0$ on a 1D numerical grid. We use a set of equally spaced points allowing us to use discrete Fourier transformations to evaluate the kinetic energy operator and the interaction term $\Phi_\sigma$. To improve accuracy we implement an axial cutoff of the $k$-space kernel $\Ut_\sigma$:  {This is obtained by Fourier transforming the  real-space interaction potential \cite{Sinha2007a} restricted to the $z$-spatial range of the grid used for the numerical calculation.} The orbital $\psi_0$ is solved using a gradient flow technique for given values of $l$ and $\eta$, thus determining a minimum energy solution of $\mathcal{E}_z$ (\ref{Ez}). An optimization scheme is used to adjust $l$ and $\eta$,   then $\psi_0$ is solved with the new parameters, and this procedure iterates until the minimum of the full energy functional (\ref{eGPE_Efun}) is found.

The 3D ground states are obtained using 3D numerical grids and discrete Fourier transforms. A cylindrically cutoff $k$-space kernel is used to improve accuracy of the $\Phi$ evaluation. The ground states are found using a conjugate gradient technique to minimize the energy functional (also see \cite{Ronen2006a,Antoine2017a}).

\begin{figure}[htbp] 
   \centering
   \includegraphics[width=3.4in]{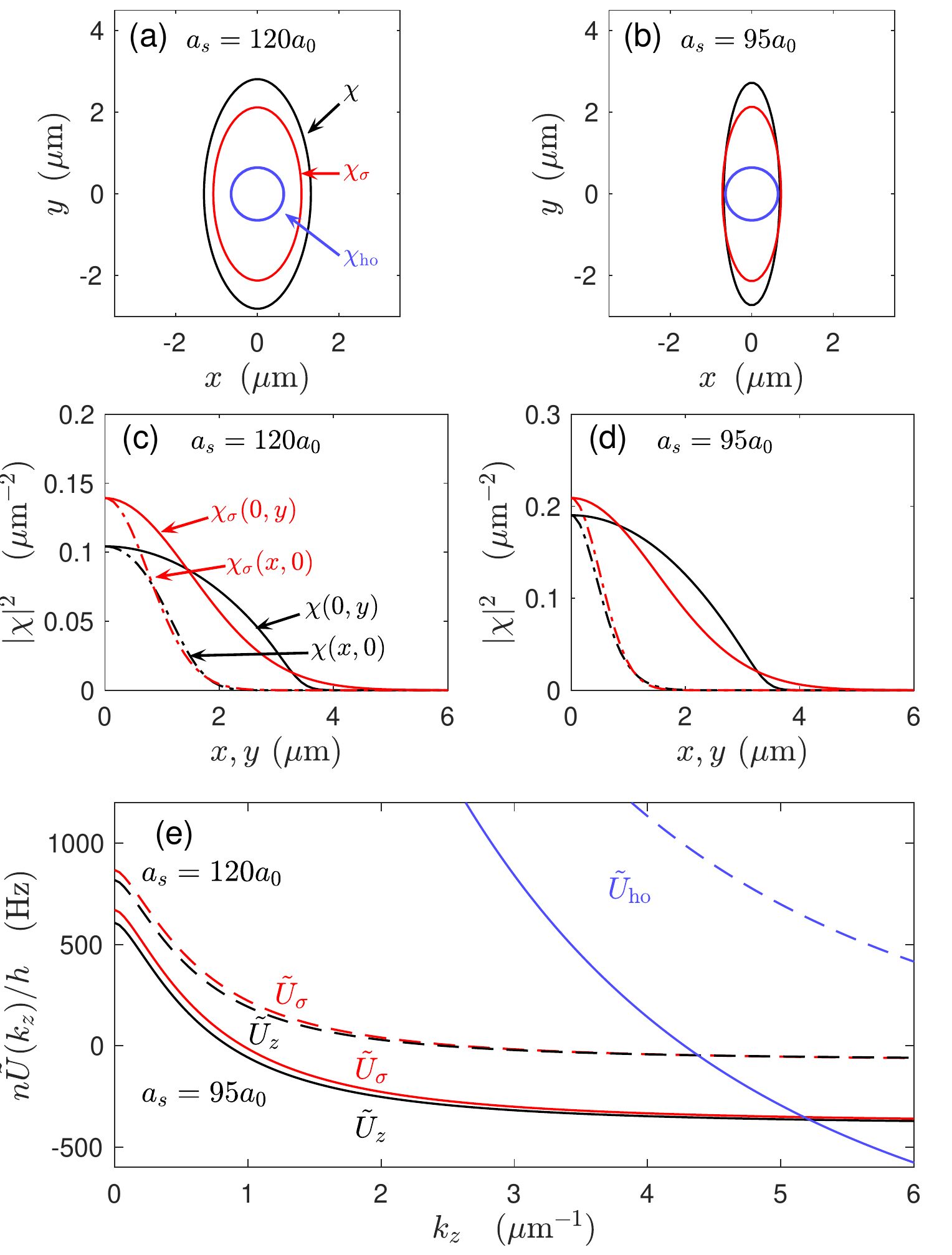}  
   \caption{ Comparison of the variational (red lines) and 3D eGPE (black lines) solutions for a uniform infinite system. The 1/e density contours of the transverse modes of the 3D eGPE $\chi$ and the variational approach $\chi_\sigma$  for (a) $a_s=120a_0$ and (b) $a_s=95a_0$. The  harmonic oscillator ground state $\chi_{\mathrm{ho}}$ is shown for reference (blue lines). In (c) and (d) we compare the transverse mode profiles along the $x$ (dash-dot) and $y$ (lines) axes for the cases given in (a) and (b), respectively. (e) The effective 1D $k$-space interaction  kernel obtained from the various transverse functions for $a_s=120a_0$ (dashed lines) and $a_s=95a_0$ (solid lines). The 3D eGPE result $\tilde{U}_{z}$ is obtained by evaluating Eq.~(\ref{Uz}) using $\chi$. The variational $\Ut_\sigma$ and the harmonic oscillator   $\tilde{U}_{\mathrm{ho}}$ results  are obtained from Eq.~(\ref{Ueta}).  Results  for $^{164}$Dy using $a_{dd}=130.8\,a_0$, with $\omega_{x,y}=2\pi\!\times\!150$Hz, $\omega_z=0$, and $n=2.5\times10^3/\mu$m.}
   \label{fig:varGPcomp}
\end{figure}

\subsection{Uniform ground states}

In Fig.~\ref{fig:varGPcomp}(a)-(d) we compare results obtained from the 3D and variational theories for the transverse density profile of a $^{164}$Dy condensate at a linear density of $n=2.5\times10^3/\mu$m for two values of $a_s$.   The lower value of  scattering length considered ($a_s=95a_0$) is close to where the roton excitation softens to zero energy and becomes dynamically unstable (see Sec.~\ref{Sec:Excitns}). For reference the harmonic oscillator ground state (i.e.~quasi-1D result) is also shown. Here we observe that both the 3D and variational eGPE solutions have a much larger transverse width than the harmonic oscillator ground state since the system parameters are outside quasi-1D regime\footnote{   {We also note that the case in Figs.~\ref{fig:varGPcomp}(b) and (d) has $na_s=12.6$, well-satisfying the requirement $na_{s}\gg 1$ established in Ref.~\cite{Edler2017a} for the quantum fluctuations to be described by the 3D formalism we use here.}}.  We also note that while the confining potential is isotropic the condensates exhibits  {magnetostriction}, i.e.~significantly elongates in the $y$-direction compared to the $x$-direction.
 
In  Fig.~\ref{fig:varGPcomp}(e) we compare the effective 1D $k$-space interaction kernel for the various theories. While the uniform ground state only depends on the value of the kernel at $k_z=0$  (\ref{Eunif}), the excitations are sensitive to its non-zero $k_z$ behaviour  (\ref{BdGUniform}). Our results show that our approximate kernel $\Ut_\sigma$ closely matches that obtained from the full 3D eGPE solution. In comparison, the quasi-1D kernel based on the harmonic oscillator ground state (see Sec.~\ref{Sec:q1D}) is a poor approximation.

\begin{figure}[htbp] 
   \centering
   \includegraphics[width=3.4in]{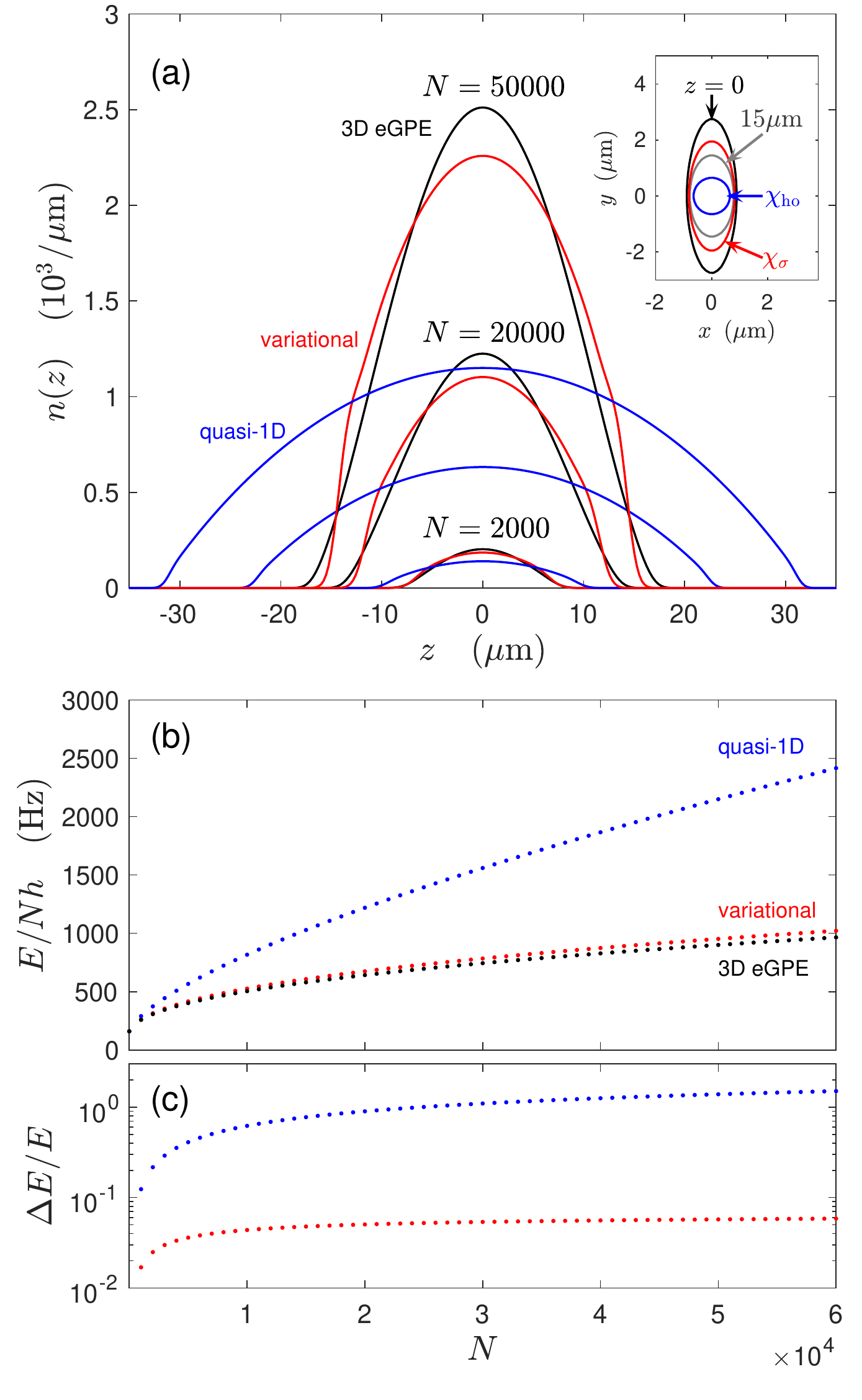}  
   \caption{Comparison of line density profiles $n(z)=\int \text{d}\bm{\rho}|\Psi(\bm{x})|^2$ and energies for a $^{164}$Dy condensate in a  trap with $\omega_{x,y}=2\pi\!\times150$Hz, $\omega_z=2\pi\times20$Hz, for $a_s=100a_0$ and various atom numbers $N$. (a) The line density along the $z$ axis calculated using the 3D eGPE (black), variational (red) and the quasi-1D (blue) theories. Inset shows the 1/e density contours (relative to the density at $\bm{\rho}=\mathbf{0}$) in the $\bm{\rho}$-plane for the various theories for $N=5\times10^4$. The contour of the 3D eGPE solution is  evaluated at $z=0$ (black) and $z=15\mu$m (grey).  (b) Energy per particle of the three theories and (c) error in the energy of the variational and the quasi-1D theories relative to the 3D eGPE results.
   }
   \label{fig:trapcomp}
\end{figure}

\subsection{Fully trapped ground states}
In Fig.~\ref{fig:trapcomp} we present results for ground states with axial confinement. The line density profiles of the solutions reveal that the variational theory is in reasonable agreement with the 3D eGPE solution, although it generally tends to have a lower peak density. Except for small atom numbers $N$, for which the interaction effects are negligible, the quasi-1D case is in poor agreement with the other theories. A significant difference between the variational and the 3D result arises because the variational solution has the separable form $\Psi(\bm{x})=\psi_0(z)\chi_\sigma(\bm{\rho})$, and thus has the same transverse profile for all $z$. At higher $N$ we can see that this not a good approximation to the 3D solution: The transverse profile at $z=0$ (where the density is highest) is more strongly affected by interactions (larger average width and anisotropy) than it is for higher values of $|z|$ [ {see inset to Fig.~\ref{fig:trapcomp}(a)}]. For the case of condensates with contact interactions an effective 1D theory has been developed that allows the transverse profile $\chi_\sigma$ to vary slowly with $z$ (see Ref.~\cite{Salasnich2002a}). Such a theory for could be developed for the dipolar case, although we do not pursue this here (also see \cite{Knight2019a}).
In  Fig.~\ref{fig:trapcomp}(b) and (c) we compare the ground state energy, observing that over the wide parameter regime considered the variational prediction for the energy is typically within a few percent of the full 3D solution.

\subsection{Uniform system excitations: Roton softening}\label{Sec:Excitns}

\begin{figure}[htbp] 
   \centering
   \includegraphics[width=3.4in]{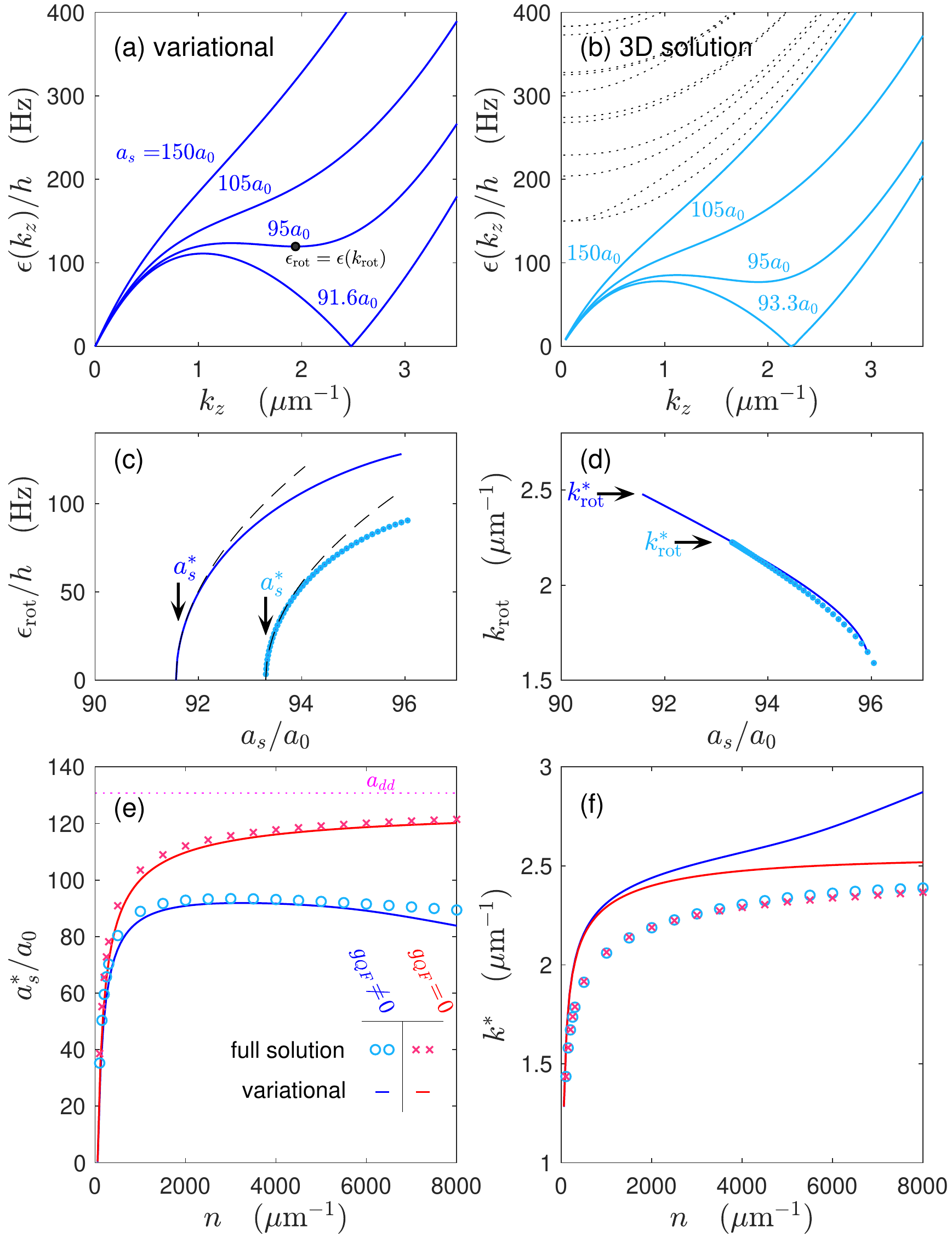}  
   \caption{Excitation dispersion relations obtained from the (a) variational (dark blue) and (b) 3D   (light blue) BdG theories for various values of $a_s$ as labelled. In (a) the black circle indicates the roton coordinates ($k_{\mathrm{rot}},\epsilon_{\mathrm{rot}}$) for $a_s=95a_0$. In (b) the higher excitations bands for $a_s=150a_0$ are shown (black dotted lines). The (c) roton energy and (d) roton wavevector as $a_s$ changes for the variational (dark blue line) and 3D (light blue dots) theories. In (c) the critical value $a^*_s$ at which the roton energy goes to zero is indicated for each theory with an arrow and the fit function $\alpha\sqrt{a_s-a^*_s}$ (with $\alpha$ a fitting parameter) is also shown (dashed lines).   In (d) the value of the critical roton wavevector ($k^*_{\mathrm{rot}}$) when $\epsilon_{\mathrm{rot}}=0$ for $a_s=a_s^*$ is indicated by an arrow for each theory.
   The roton critical values (e)  $a^*_s$ and (f) $k^*_{\mathrm{rot}}$  as the system density changes. We compare the variational (lines) and 3D (symbols) theories both including (dark blue line for variational, light blue circles for 3D) and neglecting (red line for variational, magenta crosses for 3D) the quantum fluctuation term.
Results  for $^{164}$Dy with $\omega_{x,y}=2\pi\times150$Hz, and in (a)-(d) the density is $n=2.5\times10^3/\mu$m.}
\label{fig:rotonfig}
\end{figure}

In Figs.~\ref{fig:rotonfig}(a) and (b) we compare the predictions of the variational and 3D theories for the spectrum of a uniform case as $a_s$ is varied. In these results we see that a roton (i.e.~a local minimum in the excitation dispersion relation) appears for $a_s\approx95a_0$ and lowers in energy as $a_s$ is further decreased. Our calculations predict that the roton hits zero energy at the critical value of scattering length $a^*_s$ with $a^*_s=91.6a_0$ ($a^*_s=93.3a_0$) according the variational (3D) theory for the density considered. In general we find that the variational theory predicts a lower value of $a_s^*$ than the 3D result [also see Figs.~\ref{fig:rotonfig}(c) and (e)]. For $a_s<a^*_s$ the uniform state is dynamically unstable.

Identifying the local minimum in the dispersion relation with the roton energy $\epsilon_{\mathrm{rot}}$ and wavevector $k_{\mathrm{rot}}$ [see Fig.~\ref{fig:rotonfig}(a)], we can monitor the behavior of the roton as $a_s$ varies in Figs.~\ref{fig:rotonfig}(c) and (d). The roton energy is seen to soften to zero as $(a_s-a^*_s)^{1/2}$ for $a_s$ above but close to $a^*_s$ \cite{Chomaz2018a}. The roton wave vector tends to increase as $a_s$ decreases, and obtains the value $k^*_{\mathrm{rot}}$ at the critical point $a^*_s$.   
We observe that the variational theory predicts $k^*_{\mathrm{rot}}$ to be larger than that obtained from  the 3D theory, however the $k_{\mathrm{rot}}$ values of both theories are similar at the same $a_s$ values [see Fig.~\ref{fig:rotonfig}(d)]. 
We note that our results show that the roton wave vector occurs at a value slightly higher than the inverse harmonic oscillator length along the dipole direction, i.e.~$\sqrt{m\omega_y/\hbar}=1.56/\mu $m, similar to the observations of experiments  \cite{Chomaz2018a}.

In Figs.~\ref{fig:rotonfig}(e) and (f) we examine the values of $a^*_s$ and $k^*_{\mathrm{rot}}$ for a range of system densities. We also include results without the quantum fluctuation term.  For small $n$,  the quantum fluctuations have a small effect and the theories make similar predictions\footnote{ {The results for $n\lesssim500/\mu$m on Figs.~\ref{fig:rotonfig}(e) and (f) have $na_s\lesssim1$, and the 3D treatment of quantum fluctuations is inappropriate. A quantitative treatment of this regime is outside the scope of this work.}}.  However, in general we find that $a^*_s$ is larger when the  quantum fluctuation term is neglected. We understand this arises because the quantum fluctuations effectively act as a repulsive interaction and tend to stabilize (i.e.~lift the energy) of the roton. Thus with quantum fluctuations a lower $a_s$ value is needed to destabilize the condensate. We also find that $a^*_s$ as a function of $n$  has a maximum when quantum fluctuations are included.  In contrast without quantum fluctuations $a^*_s$ monotonically increases with $n$, slowly approaching the value $a_{dd}$ in the large density limit.

\section{Conclusions and outlook}\label{Sec:Conclusions}
 
In this paper we have reported the development of a simple theory for a dipolar condensate in an elongated confining potential. Our main result is the  effective 1D variational eGPE for stationary states, and the associated BdG theory of its collective excitations. This theory is practical to solve with modest computational resources, yet provides a good quantitative description of the full 3D solution.

 In the application of our theory we have focused on the typical density, interaction and trap parameter regimes used in current experiments. For example, the rotons observed in the experiments of Chomaz \textit{et al.}~\cite{Chomaz2018a} occurred (prior to structure formation occurring) in an elongated system with linear densities in the range $2\times10^3-4\times10^3 \mu$m$^{-1}$. This regime is well-beyond where the quasi-1D approximation is valid. Using our theory we have  predicted the scattering length $a_s^*$ and roton wave vector $k_{\mathrm{rot}}^*$ at the point where roton softens to zero energy as a function of system density. Our results show that quantum fluctuations  lower the value of $a^*_s$  and cause it to have a non-monotonic dependence on density. These predictions could be investigated in future experiments and may relate to the non-monotonic dependence of the value of $a_s$ where the supersolid transition was observed [see Fig.~1(g) of Ref.~\cite{Chomaz2019a}].

Here we have restricted our focus to the regime where the system does not develop density modulations, which tends to occur at lower values of $a_s$ (e.g., after the roton softens and causes a dynamic instability). Such modulations can indicate the onset of a supersolid state, as has been observed in three recent experiments working with dipolar condensates in elongated trapping potentials.
 So far theoretical studies of the ground states and their excitations of these modulated states required large scale numerical methods for cases with (see Refs.~\cite{Tanzi2019a,Bottcher2019a,Chomaz2019a,Tanzi2019b,Guo2019a,Natale2019a}) and without  (see~\cite{Roccuzzo2019a}) axial confinement.  Our theory can treat such modulated states, but a full and systematic treatment of this is beyond the current scope and will be examined in  future work.

\begin{acknowledgments}
We acknowledge the contribution of NZ eScience Infrastructure (NeSI) high-performance computing facilities,  support from the Marsden Fund of the Royal Society of New Zealand, and valuable discussions with F.~Ferlaino and L.~Chomaz.
  
\end{acknowledgments}


\begin{thebibliography}{46}%
\makeatletter
\providecommand \@ifxundefined [1]{%
 \@ifx{#1\undefined}
}%
\providecommand \@ifnum [1]{%
 \ifnum #1\expandafter \@firstoftwo
 \else \expandafter \@secondoftwo
 \fi
}%
\providecommand \@ifx [1]{%
 \ifx #1\expandafter \@firstoftwo
 \else \expandafter \@secondoftwo
 \fi
}%
\providecommand \natexlab [1]{#1}%
\providecommand \enquote  [1]{``#1''}%
\providecommand \bibnamefont  [1]{#1}%
\providecommand \bibfnamefont [1]{#1}%
\providecommand \citenamefont [1]{#1}%
\providecommand \href@noop [0]{\@secondoftwo}%
\providecommand \href [0]{\begingroup \@sanitize@url \@href}%
\providecommand \@href[1]{\@@startlink{#1}\@@href}%
\providecommand \@@href[1]{\endgroup#1\@@endlink}%
\providecommand \@sanitize@url [0]{\catcode `\\12\catcode `\$12\catcode
  `\&12\catcode `\#12\catcode `\^12\catcode `\_12\catcode `\%12\relax}%
\providecommand \@@startlink[1]{}%
\providecommand \@@endlink[0]{}%
\providecommand \url  [0]{\begingroup\@sanitize@url \@url }%
\providecommand \@url [1]{\endgroup\@href {#1}{\urlprefix }}%
\providecommand \urlprefix  [0]{URL }%
\providecommand \Eprint [0]{\href }%
\providecommand \doibase [0]{http://dx.doi.org/}%
\providecommand \selectlanguage [0]{\@gobble}%
\providecommand \bibinfo  [0]{\@secondoftwo}%
\providecommand \bibfield  [0]{\@secondoftwo}%
\providecommand \translation [1]{[#1]}%
\providecommand \BibitemOpen [0]{}%
\providecommand \bibitemStop [0]{}%
\providecommand \bibitemNoStop [0]{.\EOS\space}%
\providecommand \EOS [0]{\spacefactor3000\relax}%
\providecommand \BibitemShut  [1]{\csname bibitem#1\endcsname}%
\let\auto@bib@innerbib\@empty
\bibitem [{\citenamefont {Griesmaier}\ \emph {et~al.}(2005)\citenamefont
  {Griesmaier}, \citenamefont {Werner}, \citenamefont {Hensler}, \citenamefont
  {Stuhler},\ and\ \citenamefont {Pfau}}]{Griesmaier2005a}%
  \BibitemOpen
  \bibfield  {author} {\bibinfo {author} {\bibfnamefont {Axel}\ \bibnamefont
  {Griesmaier}}, \bibinfo {author} {\bibfnamefont {J\"org}\ \bibnamefont
  {Werner}}, \bibinfo {author} {\bibfnamefont {Sven}\ \bibnamefont {Hensler}},
  \bibinfo {author} {\bibfnamefont {J\"urgen}\ \bibnamefont {Stuhler}}, \ and\
  \bibinfo {author} {\bibfnamefont {Tilman}\ \bibnamefont {Pfau}},\ }\bibfield
  {title} {\enquote {\bibinfo {title} {{B}ose-{E}instein condensation of
  chromium},}\ }\href {\doibase 10.1103/PhysRevLett.94.160401} {\bibfield
  {journal} {\bibinfo  {journal} {Phys. Rev. Lett.}\ }\textbf {\bibinfo
  {volume} {94}},\ \bibinfo {pages} {160401} (\bibinfo {year}
  {2005})}\BibitemShut {NoStop}%
\bibitem [{\citenamefont {Pasquiou}\ \emph {et~al.}(2011)\citenamefont
  {Pasquiou}, \citenamefont {Bismut}, \citenamefont {Mar\'echal}, \citenamefont
  {Pedri}, \citenamefont {Vernac}, \citenamefont {Gorceix},\ and\ \citenamefont
  {Laburthe-Tolra}}]{Pasquiou2011a}%
  \BibitemOpen
  \bibfield  {author} {\bibinfo {author} {\bibfnamefont {B.}~\bibnamefont
  {Pasquiou}}, \bibinfo {author} {\bibfnamefont {G.}~\bibnamefont {Bismut}},
  \bibinfo {author} {\bibfnamefont {E.}~\bibnamefont {Mar\'echal}}, \bibinfo
  {author} {\bibfnamefont {P.}~\bibnamefont {Pedri}}, \bibinfo {author}
  {\bibfnamefont {L.}~\bibnamefont {Vernac}}, \bibinfo {author} {\bibfnamefont
  {O.}~\bibnamefont {Gorceix}}, \ and\ \bibinfo {author} {\bibfnamefont
  {B.}~\bibnamefont {Laburthe-Tolra}},\ }\bibfield  {title} {\enquote {\bibinfo
  {title} {Spin relaxation and band excitation of a dipolar {B}ose-{E}instein
  condensate in 2{D} optical lattices},}\ }\href {\doibase
  10.1103/PhysRevLett.106.015301} {\bibfield  {journal} {\bibinfo  {journal}
  {Phys. Rev. Lett.}\ }\textbf {\bibinfo {volume} {106}},\ \bibinfo {pages}
  {015301} (\bibinfo {year} {2011})}\BibitemShut {NoStop}%
\bibitem [{\citenamefont {Lu}\ \emph {et~al.}(2011)\citenamefont {Lu},
  \citenamefont {Burdick}, \citenamefont {Youn},\ and\ \citenamefont
  {Lev}}]{Mingwu2011a}%
  \BibitemOpen
  \bibfield  {author} {\bibinfo {author} {\bibfnamefont {Mingwu}\ \bibnamefont
  {Lu}}, \bibinfo {author} {\bibfnamefont {Nathaniel~Q.}\ \bibnamefont
  {Burdick}}, \bibinfo {author} {\bibfnamefont {Seo~Ho}\ \bibnamefont {Youn}},
  \ and\ \bibinfo {author} {\bibfnamefont {Benjamin~L.}\ \bibnamefont {Lev}},\
  }\bibfield  {title} {\enquote {\bibinfo {title} {Strongly dipolar
  {B}ose-{E}instein condensate of dysprosium},}\ }\href {\doibase
  10.1103/PhysRevLett.107.190401} {\bibfield  {journal} {\bibinfo  {journal}
  {Phys. Rev. Lett.}\ }\textbf {\bibinfo {volume} {107}},\ \bibinfo {pages}
  {190401} (\bibinfo {year} {2011})}\BibitemShut {NoStop}%
\bibitem [{\citenamefont {Lu}\ \emph {et~al.}(2012)\citenamefont {Lu},
  \citenamefont {Burdick},\ and\ \citenamefont {Lev}}]{Lu2012a}%
  \BibitemOpen
  \bibfield  {author} {\bibinfo {author} {\bibfnamefont {Mingwu}\ \bibnamefont
  {Lu}}, \bibinfo {author} {\bibfnamefont {Nathaniel~Q.}\ \bibnamefont
  {Burdick}}, \ and\ \bibinfo {author} {\bibfnamefont {Benjamin~L.}\
  \bibnamefont {Lev}},\ }\bibfield  {title} {\enquote {\bibinfo {title}
  {Quantum degenerate dipolar {F}ermi gas},}\ }\href {\doibase
  10.1103/PhysRevLett.108.215301} {\bibfield  {journal} {\bibinfo  {journal}
  {Phys. Rev. Lett.}\ }\textbf {\bibinfo {volume} {108}},\ \bibinfo {pages}
  {215301} (\bibinfo {year} {2012})}\BibitemShut {NoStop}%
\bibitem [{\citenamefont {Aikawa}\ \emph {et~al.}(2012)\citenamefont {Aikawa},
  \citenamefont {Frisch}, \citenamefont {Mark}, \citenamefont {Baier},
  \citenamefont {Rietzler}, \citenamefont {Grimm},\ and\ \citenamefont
  {Ferlaino}}]{Aikawa2012a}%
  \BibitemOpen
  \bibfield  {author} {\bibinfo {author} {\bibfnamefont {K.}~\bibnamefont
  {Aikawa}}, \bibinfo {author} {\bibfnamefont {A.}~\bibnamefont {Frisch}},
  \bibinfo {author} {\bibfnamefont {M.}~\bibnamefont {Mark}}, \bibinfo {author}
  {\bibfnamefont {S.}~\bibnamefont {Baier}}, \bibinfo {author} {\bibfnamefont
  {A.}~\bibnamefont {Rietzler}}, \bibinfo {author} {\bibfnamefont
  {R.}~\bibnamefont {Grimm}}, \ and\ \bibinfo {author} {\bibfnamefont
  {F.}~\bibnamefont {Ferlaino}},\ }\bibfield  {title} {\enquote {\bibinfo
  {title} {{B}ose-{E}instein condensation of erbium},}\ }\href {\doibase
  10.1103/PhysRevLett.108.210401} {\bibfield  {journal} {\bibinfo  {journal}
  {Phys. Rev. Lett.}\ }\textbf {\bibinfo {volume} {108}},\ \bibinfo {pages}
  {210401} (\bibinfo {year} {2012})}\BibitemShut {NoStop}%
\bibitem [{\citenamefont {Chomaz}\ \emph {et~al.}(2018)\citenamefont {Chomaz},
  \citenamefont {van Bijnen}, \citenamefont {Petter}, \citenamefont {Faraoni},
  \citenamefont {Baier}, \citenamefont {Becher}, \citenamefont {Mark},
  \citenamefont {W{\"a}chtler}, \citenamefont {Santos},\ and\ \citenamefont
  {Ferlaino}}]{Chomaz2018a}%
  \BibitemOpen
  \bibfield  {author} {\bibinfo {author} {\bibfnamefont {L.}~\bibnamefont
  {Chomaz}}, \bibinfo {author} {\bibfnamefont {R.~M.~W.}\ \bibnamefont {van
  Bijnen}}, \bibinfo {author} {\bibfnamefont {D.}~\bibnamefont {Petter}},
  \bibinfo {author} {\bibfnamefont {G.}~\bibnamefont {Faraoni}}, \bibinfo
  {author} {\bibfnamefont {S.}~\bibnamefont {Baier}}, \bibinfo {author}
  {\bibfnamefont {J.~H.}\ \bibnamefont {Becher}}, \bibinfo {author}
  {\bibfnamefont {M.~J.}\ \bibnamefont {Mark}}, \bibinfo {author}
  {\bibfnamefont {F.}~\bibnamefont {W{\"a}chtler}}, \bibinfo {author}
  {\bibfnamefont {L.}~\bibnamefont {Santos}}, \ and\ \bibinfo {author}
  {\bibfnamefont {F.}~\bibnamefont {Ferlaino}},\ }\bibfield  {title} {\enquote
  {\bibinfo {title} {Observation of roton mode population in a dipolar quantum
  gas},}\ }\href {\doibase 10.1038/s41567-018-0054-7} {\bibfield  {journal}
  {\bibinfo  {journal} {Nat. Phys.}\ }\textbf {\bibinfo {volume} {14}},\
  \bibinfo {pages} {442} (\bibinfo {year} {2018})}\BibitemShut {NoStop}%
\bibitem [{\citenamefont {Petter}\ \emph {et~al.}(2019)\citenamefont {Petter},
  \citenamefont {Natale}, \citenamefont {van Bijnen}, \citenamefont
  {Patscheider}, \citenamefont {Mark}, \citenamefont {Chomaz},\ and\
  \citenamefont {Ferlaino}}]{Petter2018a}%
  \BibitemOpen
  \bibfield  {author} {\bibinfo {author} {\bibfnamefont {D.}~\bibnamefont
  {Petter}}, \bibinfo {author} {\bibfnamefont {G.}~\bibnamefont {Natale}},
  \bibinfo {author} {\bibfnamefont {R.~M.~W.}\ \bibnamefont {van Bijnen}},
  \bibinfo {author} {\bibfnamefont {A.}~\bibnamefont {Patscheider}}, \bibinfo
  {author} {\bibfnamefont {M.~J.}\ \bibnamefont {Mark}}, \bibinfo {author}
  {\bibfnamefont {L.}~\bibnamefont {Chomaz}}, \ and\ \bibinfo {author}
  {\bibfnamefont {F.}~\bibnamefont {Ferlaino}},\ }\bibfield  {title} {\enquote
  {\bibinfo {title} {Probing the roton excitation spectrum of a stable dipolar
  {B}ose gas},}\ }\href {\doibase 10.1103/PhysRevLett.122.183401} {\bibfield
  {journal} {\bibinfo  {journal} {Phys. Rev. Lett.}\ }\textbf {\bibinfo
  {volume} {122}},\ \bibinfo {pages} {183401} (\bibinfo {year}
  {2019})}\BibitemShut {NoStop}%
\bibitem [{\citenamefont {Landau}(1941)}]{Landau1941a}%
  \BibitemOpen
  \bibfield  {author} {\bibinfo {author} {\bibfnamefont {L.~D.}\ \bibnamefont
  {Landau}},\ }\bibfield  {title} {\enquote {\bibinfo {title} {The theory of
  superfluidity of helium {II}},}\ }\href@noop {} {\bibfield  {journal}
  {\bibinfo  {journal} {J. Phys. (Mosc.)}\ }\textbf {\bibinfo {volume} {5}},\
  \bibinfo {pages} {71} (\bibinfo {year} {1941})}\BibitemShut {NoStop}%
\bibitem [{\citenamefont {Santos}\ \emph {et~al.}(2003)\citenamefont {Santos},
  \citenamefont {Shlyapnikov},\ and\ \citenamefont {Lewenstein}}]{Santos2003a}%
  \BibitemOpen
  \bibfield  {author} {\bibinfo {author} {\bibfnamefont {L.}~\bibnamefont
  {Santos}}, \bibinfo {author} {\bibfnamefont {G.~V.}\ \bibnamefont
  {Shlyapnikov}}, \ and\ \bibinfo {author} {\bibfnamefont {M.}~\bibnamefont
  {Lewenstein}},\ }\bibfield  {title} {\enquote {\bibinfo {title} {Roton-maxon
  spectrum and stability of trapped dipolar {B}ose-{E}instein condensates},}\
  }\href {\doibase 10.1103/PhysRevLett.90.250403} {\bibfield  {journal}
  {\bibinfo  {journal} {Phys. Rev. Lett.}\ }\textbf {\bibinfo {volume} {90}},\
  \bibinfo {pages} {250403} (\bibinfo {year} {2003})}\BibitemShut {NoStop}%
\bibitem [{\citenamefont {Ronen}\ \emph {et~al.}(2007)\citenamefont {Ronen},
  \citenamefont {Bortolotti},\ and\ \citenamefont {Bohn}}]{Ronen2007a}%
  \BibitemOpen
  \bibfield  {author} {\bibinfo {author} {\bibfnamefont {Shai}\ \bibnamefont
  {Ronen}}, \bibinfo {author} {\bibfnamefont {Daniele C.~E.}\ \bibnamefont
  {Bortolotti}}, \ and\ \bibinfo {author} {\bibfnamefont {John~L.}\
  \bibnamefont {Bohn}},\ }\bibfield  {title} {\enquote {\bibinfo {title}
  {Radial and angular rotons in trapped dipolar gases},}\ }\href {\doibase
  10.1103/PhysRevLett.98.030406} {\bibfield  {journal} {\bibinfo  {journal}
  {Phys. Rev. Lett.}\ }\textbf {\bibinfo {volume} {98}},\ \bibinfo {eid}
  {030406} (\bibinfo {year} {2007})}\BibitemShut {NoStop}%
\bibitem [{\citenamefont {Blakie}\ \emph {et~al.}(2012)\citenamefont {Blakie},
  \citenamefont {Baillie},\ and\ \citenamefont {Bisset}}]{Blakie2012a}%
  \BibitemOpen
  \bibfield  {author} {\bibinfo {author} {\bibfnamefont {P.~B.}\ \bibnamefont
  {Blakie}}, \bibinfo {author} {\bibfnamefont {D.}~\bibnamefont {Baillie}}, \
  and\ \bibinfo {author} {\bibfnamefont {R.~N.}\ \bibnamefont {Bisset}},\
  }\bibfield  {title} {\enquote {\bibinfo {title} {Roton spectroscopy in a
  harmonically trapped dipolar {B}ose-{E}instein condensate},}\ }\href
  {\doibase 10.1103/PhysRevA.86.021604} {\bibfield  {journal} {\bibinfo
  {journal} {Phys. Rev. A}\ }\textbf {\bibinfo {volume} {86}},\ \bibinfo
  {pages} {021604} (\bibinfo {year} {2012})}\BibitemShut {NoStop}%
\bibitem [{\citenamefont {Corson}\ \emph
  {et~al.}(2013{\natexlab{a}})\citenamefont {Corson}, \citenamefont {Wilson},\
  and\ \citenamefont {Bohn}}]{Corson2013a}%
  \BibitemOpen
  \bibfield  {author} {\bibinfo {author} {\bibfnamefont {John~P.}\ \bibnamefont
  {Corson}}, \bibinfo {author} {\bibfnamefont {Ryan~M.}\ \bibnamefont
  {Wilson}}, \ and\ \bibinfo {author} {\bibfnamefont {John~L.}\ \bibnamefont
  {Bohn}},\ }\bibfield  {title} {\enquote {\bibinfo {title} {Stability
  spectroscopy of rotons in a dipolar {B}ose gas},}\ }\href {\doibase
  10.1103/PhysRevA.87.051605} {\bibfield  {journal} {\bibinfo  {journal} {Phys.
  Rev. A}\ }\textbf {\bibinfo {volume} {87}},\ \bibinfo {pages} {051605}
  (\bibinfo {year} {2013}{\natexlab{a}})}\BibitemShut {NoStop}%
\bibitem [{\citenamefont {Corson}\ \emph
  {et~al.}(2013{\natexlab{b}})\citenamefont {Corson}, \citenamefont {Wilson},\
  and\ \citenamefont {Bohn}}]{Corson2013b}%
  \BibitemOpen
  \bibfield  {author} {\bibinfo {author} {\bibfnamefont {John~P.}\ \bibnamefont
  {Corson}}, \bibinfo {author} {\bibfnamefont {Ryan~M.}\ \bibnamefont
  {Wilson}}, \ and\ \bibinfo {author} {\bibfnamefont {John~L.}\ \bibnamefont
  {Bohn}},\ }\bibfield  {title} {\enquote {\bibinfo {title} {Geometric
  stability spectra of dipolar {B}ose gases in tunable optical lattices},}\
  }\href {\doibase 10.1103/PhysRevA.88.013614} {\bibfield  {journal} {\bibinfo
  {journal} {Phys. Rev. A}\ }\textbf {\bibinfo {volume} {88}},\ \bibinfo
  {pages} {013614} (\bibinfo {year} {2013}{\natexlab{b}})}\BibitemShut
  {NoStop}%
\bibitem [{\citenamefont {Jona-Lasinio}\ \emph {et~al.}(2013)\citenamefont
  {Jona-Lasinio}, \citenamefont {\L{}akomy},\ and\ \citenamefont
  {Santos}}]{JonaLasinio2013b}%
  \BibitemOpen
  \bibfield  {author} {\bibinfo {author} {\bibfnamefont {M.}~\bibnamefont
  {Jona-Lasinio}}, \bibinfo {author} {\bibfnamefont {K.}~\bibnamefont
  {\L{}akomy}}, \ and\ \bibinfo {author} {\bibfnamefont {L.}~\bibnamefont
  {Santos}},\ }\bibfield  {title} {\enquote {\bibinfo {title} {Time-of-flight
  roton spectroscopy in dipolar {B}ose-{E}instein condensates},}\ }\href
  {\doibase 10.1103/PhysRevA.88.025603} {\bibfield  {journal} {\bibinfo
  {journal} {Phys. Rev. A}\ }\textbf {\bibinfo {volume} {88}},\ \bibinfo
  {pages} {025603} (\bibinfo {year} {2013})}\BibitemShut {NoStop}%
\bibitem [{\citenamefont {Bisset}\ and\ \citenamefont
  {Blakie}(2013)}]{Bisset2013a}%
  \BibitemOpen
  \bibfield  {author} {\bibinfo {author} {\bibfnamefont {R.~N.}\ \bibnamefont
  {Bisset}}\ and\ \bibinfo {author} {\bibfnamefont {P.~B.}\ \bibnamefont
  {Blakie}},\ }\bibfield  {title} {\enquote {\bibinfo {title} {Fingerprinting
  rotons in a dipolar condensate: Super-poissonian peak in the atom-number
  fluctuations},}\ }\href {\doibase 10.1103/PhysRevLett.110.265302} {\bibfield
  {journal} {\bibinfo  {journal} {Phys. Rev. Lett.}\ }\textbf {\bibinfo
  {volume} {110}},\ \bibinfo {pages} {265302} (\bibinfo {year}
  {2013})}\BibitemShut {NoStop}%
\bibitem [{\citenamefont {Baillie}\ and\ \citenamefont
  {Blakie}(2015)}]{Baillie2015b}%
  \BibitemOpen
  \bibfield  {author} {\bibinfo {author} {\bibfnamefont {D}~\bibnamefont
  {Baillie}}\ and\ \bibinfo {author} {\bibfnamefont {P~B}\ \bibnamefont
  {Blakie}},\ }\bibfield  {title} {\enquote {\bibinfo {title} {A general theory
  of flattened dipolar condensates},}\ }\href {\doibase
  10.1088/1367-2630/17/3/033028} {\bibfield  {journal} {\bibinfo  {journal}
  {New J. Phys}\ }\textbf {\bibinfo {volume} {17}},\ \bibinfo {pages} {033028}
  (\bibinfo {year} {2015})}\BibitemShut {NoStop}%
\bibitem [{\citenamefont {Roccuzzo}\ and\ \citenamefont
  {Ancilotto}(2019)}]{Roccuzzo2019a}%
  \BibitemOpen
  \bibfield  {author} {\bibinfo {author} {\bibfnamefont {Santo~Maria}\
  \bibnamefont {Roccuzzo}}\ and\ \bibinfo {author} {\bibfnamefont {Francesco}\
  \bibnamefont {Ancilotto}},\ }\bibfield  {title} {\enquote {\bibinfo {title}
  {Supersolid behavior of a dipolar {B}ose-{E}instein condensate confined in a
  tube},}\ }\href {\doibase 10.1103/PhysRevA.99.041601} {\bibfield  {journal}
  {\bibinfo  {journal} {Phys. Rev. A}\ }\textbf {\bibinfo {volume} {99}},\
  \bibinfo {pages} {041601} (\bibinfo {year} {2019})}\BibitemShut {NoStop}%
\bibitem [{\citenamefont {Kadau}\ \emph {et~al.}(2016)\citenamefont {Kadau},
  \citenamefont {Schmitt}, \citenamefont {Wenzel}, \citenamefont {Wink},
  \citenamefont {Maier}, \citenamefont {Ferrier-Barbut},\ and\ \citenamefont
  {Pfau}}]{Kadau2016a}%
  \BibitemOpen
  \bibfield  {author} {\bibinfo {author} {\bibfnamefont {Holger}\ \bibnamefont
  {Kadau}}, \bibinfo {author} {\bibfnamefont {Matthias}\ \bibnamefont
  {Schmitt}}, \bibinfo {author} {\bibfnamefont {Matthias}\ \bibnamefont
  {Wenzel}}, \bibinfo {author} {\bibfnamefont {Clarissa}\ \bibnamefont {Wink}},
  \bibinfo {author} {\bibfnamefont {Thomas}\ \bibnamefont {Maier}}, \bibinfo
  {author} {\bibfnamefont {Igor}\ \bibnamefont {Ferrier-Barbut}}, \ and\
  \bibinfo {author} {\bibfnamefont {Tilman}\ \bibnamefont {Pfau}},\ }\bibfield
  {title} {\enquote {\bibinfo {title} {Observing the {R}osensweig instability
  of a quantum ferrofluid},}\ }\href {http://dx.doi.org/10.1038/nature16485}
  {\bibfield  {journal} {\bibinfo  {journal} {Nature}\ }\textbf {\bibinfo
  {volume} {530}},\ \bibinfo {pages} {194--197} (\bibinfo {year}
  {2016})}\BibitemShut {NoStop}%
\bibitem [{\citenamefont {Ferrier-Barbut}\ \emph {et~al.}(2016)\citenamefont
  {Ferrier-Barbut}, \citenamefont {Kadau}, \citenamefont {Schmitt},
  \citenamefont {Wenzel},\ and\ \citenamefont {Pfau}}]{Ferrier-Barbut2016a}%
  \BibitemOpen
  \bibfield  {author} {\bibinfo {author} {\bibfnamefont {Igor}\ \bibnamefont
  {Ferrier-Barbut}}, \bibinfo {author} {\bibfnamefont {Holger}\ \bibnamefont
  {Kadau}}, \bibinfo {author} {\bibfnamefont {Matthias}\ \bibnamefont
  {Schmitt}}, \bibinfo {author} {\bibfnamefont {Matthias}\ \bibnamefont
  {Wenzel}}, \ and\ \bibinfo {author} {\bibfnamefont {Tilman}\ \bibnamefont
  {Pfau}},\ }\bibfield  {title} {\enquote {\bibinfo {title} {Observation of
  quantum droplets in a strongly dipolar {B}ose gas},}\ }\href {\doibase
  10.1103/PhysRevLett.116.215301} {\bibfield  {journal} {\bibinfo  {journal}
  {Phys. Rev. Lett.}\ }\textbf {\bibinfo {volume} {116}},\ \bibinfo {pages}
  {215301} (\bibinfo {year} {2016})}\BibitemShut {NoStop}%
\bibitem [{\citenamefont {Bisset}\ \emph {et~al.}(2016)\citenamefont {Bisset},
  \citenamefont {Wilson}, \citenamefont {Baillie},\ and\ \citenamefont
  {Blakie}}]{Bisset2016a}%
  \BibitemOpen
  \bibfield  {author} {\bibinfo {author} {\bibfnamefont {R.~N.}\ \bibnamefont
  {Bisset}}, \bibinfo {author} {\bibfnamefont {R.~M.}\ \bibnamefont {Wilson}},
  \bibinfo {author} {\bibfnamefont {D.}~\bibnamefont {Baillie}}, \ and\
  \bibinfo {author} {\bibfnamefont {P.~B.}\ \bibnamefont {Blakie}},\ }\bibfield
   {title} {\enquote {\bibinfo {title} {Ground-state phase diagram of a dipolar
  condensate with quantum fluctuations},}\ }\href {\doibase
  10.1103/PhysRevA.94.033619} {\bibfield  {journal} {\bibinfo  {journal} {Phys.
  Rev. A}\ }\textbf {\bibinfo {volume} {94}},\ \bibinfo {pages} {033619}
  (\bibinfo {year} {2016})}\BibitemShut {NoStop}%
\bibitem [{\citenamefont {Chomaz}\ \emph {et~al.}(2016)\citenamefont {Chomaz},
  \citenamefont {Baier}, \citenamefont {Petter}, \citenamefont {Mark},
  \citenamefont {W\"achtler}, \citenamefont {Santos},\ and\ \citenamefont
  {Ferlaino}}]{Chomaz2016a}%
  \BibitemOpen
  \bibfield  {author} {\bibinfo {author} {\bibfnamefont {L.}~\bibnamefont
  {Chomaz}}, \bibinfo {author} {\bibfnamefont {S.}~\bibnamefont {Baier}},
  \bibinfo {author} {\bibfnamefont {D.}~\bibnamefont {Petter}}, \bibinfo
  {author} {\bibfnamefont {M.~J.}\ \bibnamefont {Mark}}, \bibinfo {author}
  {\bibfnamefont {F.}~\bibnamefont {W\"achtler}}, \bibinfo {author}
  {\bibfnamefont {L.}~\bibnamefont {Santos}}, \ and\ \bibinfo {author}
  {\bibfnamefont {F.}~\bibnamefont {Ferlaino}},\ }\bibfield  {title} {\enquote
  {\bibinfo {title} {Quantum-fluctuation-driven crossover from a dilute
  {B}ose-{E}instein condensate to a macrodroplet in a dipolar quantum fluid},}\
  }\href {\doibase 10.1103/PhysRevX.6.041039} {\bibfield  {journal} {\bibinfo
  {journal} {Phys. Rev. X}\ }\textbf {\bibinfo {volume} {6}},\ \bibinfo {pages}
  {041039} (\bibinfo {year} {2016})}\BibitemShut {NoStop}%
\bibitem [{\citenamefont {Schmitt}\ \emph {et~al.}(2016)\citenamefont
  {Schmitt}, \citenamefont {Wenzel}, \citenamefont {B{\"o}ttcher},
  \citenamefont {Ferrier-Barbut},\ and\ \citenamefont {Pfau}}]{Schmitt2016a}%
  \BibitemOpen
  \bibfield  {author} {\bibinfo {author} {\bibfnamefont {Matthias}\
  \bibnamefont {Schmitt}}, \bibinfo {author} {\bibfnamefont {Matthias}\
  \bibnamefont {Wenzel}}, \bibinfo {author} {\bibfnamefont {Fabian}\
  \bibnamefont {B{\"o}ttcher}}, \bibinfo {author} {\bibfnamefont {Igor}\
  \bibnamefont {Ferrier-Barbut}}, \ and\ \bibinfo {author} {\bibfnamefont
  {Tilman}\ \bibnamefont {Pfau}},\ }\bibfield  {title} {\enquote {\bibinfo
  {title} {Self-bound droplets of a dilute magnetic quantum liquid},}\ }\href
  {http://dx.doi.org/10.1038/nature20126} {\bibfield  {journal} {\bibinfo
  {journal} {Nature}\ }\textbf {\bibinfo {volume} {539}},\ \bibinfo {pages}
  {259} (\bibinfo {year} {2016})}\BibitemShut {NoStop}%
\bibitem [{\citenamefont {B\"ottcher}\ \emph
  {et~al.}(2019{\natexlab{a}})\citenamefont {B\"ottcher}, \citenamefont
  {Wenzel}, \citenamefont {Schmidt}, \citenamefont {Guo}, \citenamefont
  {Langen}, \citenamefont {Ferrier-Barbut}, \citenamefont {Pfau}, \citenamefont
  {Bomb\'{\i}n}, \citenamefont {S\'anchez-Baena}, \citenamefont {Boronat},\
  and\ \citenamefont {Mazzanti}}]{Bottcher2019b}%
  \BibitemOpen
  \bibfield  {author} {\bibinfo {author} {\bibfnamefont {Fabian}\ \bibnamefont
  {B\"ottcher}}, \bibinfo {author} {\bibfnamefont {Matthias}\ \bibnamefont
  {Wenzel}}, \bibinfo {author} {\bibfnamefont {Jan-Niklas}\ \bibnamefont
  {Schmidt}}, \bibinfo {author} {\bibfnamefont {Mingyang}\ \bibnamefont {Guo}},
  \bibinfo {author} {\bibfnamefont {Tim}\ \bibnamefont {Langen}}, \bibinfo
  {author} {\bibfnamefont {Igor}\ \bibnamefont {Ferrier-Barbut}}, \bibinfo
  {author} {\bibfnamefont {Tilman}\ \bibnamefont {Pfau}}, \bibinfo {author}
  {\bibfnamefont {Ra\'ul}\ \bibnamefont {Bomb\'{\i}n}}, \bibinfo {author}
  {\bibfnamefont {Joan}\ \bibnamefont {S\'anchez-Baena}}, \bibinfo {author}
  {\bibfnamefont {Jordi}\ \bibnamefont {Boronat}}, \ and\ \bibinfo {author}
  {\bibfnamefont {Ferran}\ \bibnamefont {Mazzanti}},\ }\bibfield  {title}
  {\enquote {\bibinfo {title} {Dilute dipolar quantum droplets beyond the
  extended {G}ross-{P}itaevskii equation},}\ }\href {\doibase
  10.1103/PhysRevResearch.1.033088} {\bibfield  {journal} {\bibinfo  {journal}
  {Phys. Rev. Research}\ }\textbf {\bibinfo {volume} {1}},\ \bibinfo {pages}
  {033088} (\bibinfo {year} {2019}{\natexlab{a}})}\BibitemShut {NoStop}%
\bibitem [{\citenamefont {B\"ottcher}\ \emph
  {et~al.}(2019{\natexlab{b}})\citenamefont {B\"ottcher}, \citenamefont
  {Schmidt}, \citenamefont {Wenzel}, \citenamefont {Hertkorn}, \citenamefont
  {Guo}, \citenamefont {Langen},\ and\ \citenamefont {Pfau}}]{Bottcher2019a}%
  \BibitemOpen
  \bibfield  {author} {\bibinfo {author} {\bibfnamefont {Fabian}\ \bibnamefont
  {B\"ottcher}}, \bibinfo {author} {\bibfnamefont {Jan-Niklas}\ \bibnamefont
  {Schmidt}}, \bibinfo {author} {\bibfnamefont {Matthias}\ \bibnamefont
  {Wenzel}}, \bibinfo {author} {\bibfnamefont {Jens}\ \bibnamefont {Hertkorn}},
  \bibinfo {author} {\bibfnamefont {Mingyang}\ \bibnamefont {Guo}}, \bibinfo
  {author} {\bibfnamefont {Tim}\ \bibnamefont {Langen}}, \ and\ \bibinfo
  {author} {\bibfnamefont {Tilman}\ \bibnamefont {Pfau}},\ }\bibfield  {title}
  {\enquote {\bibinfo {title} {Transient supersolid properties in an array of
  dipolar quantum droplets},}\ }\href {\doibase 10.1103/PhysRevX.9.011051}
  {\bibfield  {journal} {\bibinfo  {journal} {Phys. Rev. X}\ }\textbf {\bibinfo
  {volume} {9}},\ \bibinfo {pages} {011051} (\bibinfo {year}
  {2019}{\natexlab{b}})}\BibitemShut {NoStop}%
\bibitem [{\citenamefont {Tanzi}\ \emph
  {et~al.}(2019{\natexlab{a}})\citenamefont {Tanzi}, \citenamefont {Lucioni},
  \citenamefont {Fam\`a}, \citenamefont {Catani}, \citenamefont {Fioretti},
  \citenamefont {Gabbanini}, \citenamefont {Bisset}, \citenamefont {Santos},\
  and\ \citenamefont {Modugno}}]{Tanzi2019a}%
  \BibitemOpen
  \bibfield  {author} {\bibinfo {author} {\bibfnamefont {L.}~\bibnamefont
  {Tanzi}}, \bibinfo {author} {\bibfnamefont {E.}~\bibnamefont {Lucioni}},
  \bibinfo {author} {\bibfnamefont {F.}~\bibnamefont {Fam\`a}}, \bibinfo
  {author} {\bibfnamefont {J.}~\bibnamefont {Catani}}, \bibinfo {author}
  {\bibfnamefont {A.}~\bibnamefont {Fioretti}}, \bibinfo {author}
  {\bibfnamefont {C.}~\bibnamefont {Gabbanini}}, \bibinfo {author}
  {\bibfnamefont {R.~N.}\ \bibnamefont {Bisset}}, \bibinfo {author}
  {\bibfnamefont {L.}~\bibnamefont {Santos}}, \ and\ \bibinfo {author}
  {\bibfnamefont {G.}~\bibnamefont {Modugno}},\ }\bibfield  {title} {\enquote
  {\bibinfo {title} {Observation of a dipolar quantum gas with metastable
  supersolid properties},}\ }\href {\doibase 10.1103/PhysRevLett.122.130405}
  {\bibfield  {journal} {\bibinfo  {journal} {Phys. Rev. Lett.}\ }\textbf
  {\bibinfo {volume} {122}},\ \bibinfo {pages} {130405} (\bibinfo {year}
  {2019}{\natexlab{a}})}\BibitemShut {NoStop}%
\bibitem [{\citenamefont {W\"achtler}\ and\ \citenamefont
  {Santos}(2016)}]{Wachtler2016a}%
  \BibitemOpen
  \bibfield  {author} {\bibinfo {author} {\bibfnamefont {F.}~\bibnamefont
  {W\"achtler}}\ and\ \bibinfo {author} {\bibfnamefont {L.}~\bibnamefont
  {Santos}},\ }\bibfield  {title} {\enquote {\bibinfo {title} {Quantum
  filaments in dipolar {B}ose-{E}instein condensates},}\ }\href {\doibase
  10.1103/PhysRevA.93.061603} {\bibfield  {journal} {\bibinfo  {journal} {Phys.
  Rev. A}\ }\textbf {\bibinfo {volume} {93}},\ \bibinfo {pages} {061603(R)}
  (\bibinfo {year} {2016})}\BibitemShut {NoStop}%
\bibitem [{\citenamefont {Baillie}\ \emph {et~al.}(2017)\citenamefont
  {Baillie}, \citenamefont {Wilson},\ and\ \citenamefont
  {Blakie}}]{Baillie2017a}%
  \BibitemOpen
  \bibfield  {author} {\bibinfo {author} {\bibfnamefont {D.}~\bibnamefont
  {Baillie}}, \bibinfo {author} {\bibfnamefont {R.~M.}\ \bibnamefont {Wilson}},
  \ and\ \bibinfo {author} {\bibfnamefont {P.~B.}\ \bibnamefont {Blakie}},\
  }\bibfield  {title} {\enquote {\bibinfo {title} {Collective excitations of
  self-bound droplets of a dipolar quantum fluid},}\ }\href {\doibase
  10.1103/PhysRevLett.119.255302} {\bibfield  {journal} {\bibinfo  {journal}
  {Phys. Rev. Lett.}\ }\textbf {\bibinfo {volume} {119}},\ \bibinfo {pages}
  {255302} (\bibinfo {year} {2017})}\BibitemShut {NoStop}%
\bibitem [{\citenamefont {Lee}\ \emph {et~al.}(2018)\citenamefont {Lee},
  \citenamefont {Baillie}, \citenamefont {Bisset},\ and\ \citenamefont
  {Blakie}}]{Lee2018a}%
  \BibitemOpen
  \bibfield  {author} {\bibinfo {author} {\bibfnamefont {Au-Chen}\ \bibnamefont
  {Lee}}, \bibinfo {author} {\bibfnamefont {D.}~\bibnamefont {Baillie}},
  \bibinfo {author} {\bibfnamefont {R.~N.}\ \bibnamefont {Bisset}}, \ and\
  \bibinfo {author} {\bibfnamefont {P.~B.}\ \bibnamefont {Blakie}},\ }\bibfield
   {title} {\enquote {\bibinfo {title} {Excitations of a vortex line in an
  elongated dipolar condensate},}\ }\href {\doibase 10.1103/PhysRevA.98.063620}
  {\bibfield  {journal} {\bibinfo  {journal} {Phys. Rev. A}\ }\textbf {\bibinfo
  {volume} {98}},\ \bibinfo {pages} {063620} (\bibinfo {year}
  {2018})}\BibitemShut {NoStop}%
\bibitem [{\citenamefont {Salasnich}\ \emph {et~al.}(2002)\citenamefont
  {Salasnich}, \citenamefont {Parola},\ and\ \citenamefont
  {Reatto}}]{Salasnich2002a}%
  \BibitemOpen
  \bibfield  {author} {\bibinfo {author} {\bibfnamefont {L.}~\bibnamefont
  {Salasnich}}, \bibinfo {author} {\bibfnamefont {A.}~\bibnamefont {Parola}}, \
  and\ \bibinfo {author} {\bibfnamefont {L.}~\bibnamefont {Reatto}},\
  }\bibfield  {title} {\enquote {\bibinfo {title} {Effective wave equations for
  the dynamics of cigar-shaped and disk-shaped {B}ose condensates},}\ }\href
  {\doibase 10.1103/PhysRevA.65.043614} {\bibfield  {journal} {\bibinfo
  {journal} {Phys. Rev. A}\ }\textbf {\bibinfo {volume} {65}},\ \bibinfo
  {pages} {043614} (\bibinfo {year} {2002})}\BibitemShut {NoStop}%
\bibitem [{\citenamefont {Edler}\ \emph {et~al.}(2017)\citenamefont {Edler},
  \citenamefont {Mishra}, \citenamefont {W\"achtler}, \citenamefont {Nath},
  \citenamefont {Sinha},\ and\ \citenamefont {Santos}}]{Edler2017a}%
  \BibitemOpen
  \bibfield  {author} {\bibinfo {author} {\bibfnamefont {D.}~\bibnamefont
  {Edler}}, \bibinfo {author} {\bibfnamefont {C.}~\bibnamefont {Mishra}},
  \bibinfo {author} {\bibfnamefont {F.}~\bibnamefont {W\"achtler}}, \bibinfo
  {author} {\bibfnamefont {R.}~\bibnamefont {Nath}}, \bibinfo {author}
  {\bibfnamefont {S.}~\bibnamefont {Sinha}}, \ and\ \bibinfo {author}
  {\bibfnamefont {L.}~\bibnamefont {Santos}},\ }\bibfield  {title} {\enquote
  {\bibinfo {title} {Quantum fluctuations in quasi-one-dimensional dipolar
  bose-einstein condensates},}\ }\href {\doibase
  10.1103/PhysRevLett.119.050403} {\bibfield  {journal} {\bibinfo  {journal}
  {Phys. Rev. Lett.}\ }\textbf {\bibinfo {volume} {119}},\ \bibinfo {pages}
  {050403} (\bibinfo {year} {2017})}\BibitemShut {NoStop}%
\bibitem [{\citenamefont {Lima}\ and\ \citenamefont
  {Pelster}(2011)}]{Lima2011a}%
  \BibitemOpen
  \bibfield  {author} {\bibinfo {author} {\bibfnamefont {Aristeu R.~P.}\
  \bibnamefont {Lima}}\ and\ \bibinfo {author} {\bibfnamefont {Axel}\
  \bibnamefont {Pelster}},\ }\bibfield  {title} {\enquote {\bibinfo {title}
  {Quantum fluctuations in dipolar {B}ose gases},}\ }\href {\doibase
  10.1103/PhysRevA.84.041604} {\bibfield  {journal} {\bibinfo  {journal} {Phys.
  Rev. A}\ }\textbf {\bibinfo {volume} {84}},\ \bibinfo {pages} {041604}
  (\bibinfo {year} {2011})}\BibitemShut {NoStop}%
\bibitem [{\citenamefont {Sinha}\ and\ \citenamefont
  {Santos}(2007)}]{Sinha2007a}%
  \BibitemOpen
  \bibfield  {author} {\bibinfo {author} {\bibfnamefont {S.}~\bibnamefont
  {Sinha}}\ and\ \bibinfo {author} {\bibfnamefont {L.}~\bibnamefont {Santos}},\
  }\bibfield  {title} {\enquote {\bibinfo {title} {Cold dipolar gases in
  quasi-one-dimensional geometries},}\ }\href {\doibase
  10.1103/PhysRevLett.99.140406} {\bibfield  {journal} {\bibinfo  {journal}
  {Phys. Rev. Lett.}\ }\textbf {\bibinfo {volume} {99}},\ \bibinfo {pages}
  {140406} (\bibinfo {year} {2007})}\BibitemShut {NoStop}%
\bibitem [{\citenamefont {Deuretzbacher}\ \emph {et~al.}(2010)\citenamefont
  {Deuretzbacher}, \citenamefont {Cremon},\ and\ \citenamefont
  {Reimann}}]{Deuretzbacher2010a}%
  \BibitemOpen
  \bibfield  {author} {\bibinfo {author} {\bibfnamefont {F.}~\bibnamefont
  {Deuretzbacher}}, \bibinfo {author} {\bibfnamefont {J.~C.}\ \bibnamefont
  {Cremon}}, \ and\ \bibinfo {author} {\bibfnamefont {S.~M.}\ \bibnamefont
  {Reimann}},\ }\bibfield  {title} {\enquote {\bibinfo {title} {Ground-state
  properties of few dipolar bosons in a quasi-one-dimensional harmonic trap},}\
  }\href {\doibase 10.1103/PhysRevA.81.063616} {\bibfield  {journal} {\bibinfo
  {journal} {Phys. Rev. A}\ }\textbf {\bibinfo {volume} {81}},\ \bibinfo
  {pages} {063616} (\bibinfo {year} {2010})}\BibitemShut {NoStop}%
\bibitem [{\citenamefont {Deuretzbacher}\ \emph {et~al.}(2013)\citenamefont
  {Deuretzbacher}, \citenamefont {Cremon},\ and\ \citenamefont
  {Reimann}}]{Deuretzbacher2013a}%
  \BibitemOpen
  \bibfield  {author} {\bibinfo {author} {\bibfnamefont {F.}~\bibnamefont
  {Deuretzbacher}}, \bibinfo {author} {\bibfnamefont {J.~C.}\ \bibnamefont
  {Cremon}}, \ and\ \bibinfo {author} {\bibfnamefont {S.~M.}\ \bibnamefont
  {Reimann}},\ }\bibfield  {title} {\enquote {\bibinfo {title} {Erratum:
  Ground-state properties of few dipolar bosons in a quasi-one-dimensional
  harmonic trap},}\ }\href {\doibase 10.1103/PhysRevA.87.039903} {\bibfield
  {journal} {\bibinfo  {journal} {Phys. Rev. A}\ }\textbf {\bibinfo {volume}
  {87}},\ \bibinfo {pages} {039903} (\bibinfo {year} {2013})}\BibitemShut
  {NoStop}%
\bibitem [{\citenamefont {Giovanazzi}\ and\ \citenamefont
  {O'Dell}(2004)}]{Giovanazzi2004a}%
  \BibitemOpen
  \bibfield  {author} {\bibinfo {author} {\bibfnamefont {S.}~\bibnamefont
  {Giovanazzi}}\ and\ \bibinfo {author} {\bibfnamefont {D.H.J.}\ \bibnamefont
  {O'Dell}},\ }\bibfield  {title} {\enquote {\bibinfo {title} {Instabilities
  and the roton spectrum of a quasi-{1D} {B}ose-{E}instein condensed gas with
  dipole-dipole interactions},}\ }\href {\doibase 10.1140/epjd/e2004-00146-7}
  {\bibfield  {journal} {\bibinfo  {journal} {Eur. Phys. J. D}\ }\textbf
  {\bibinfo {volume} {31}},\ \bibinfo {pages} {439--445} (\bibinfo {year}
  {2004})}\BibitemShut {NoStop}%
\bibitem [{\citenamefont {Ronen}\ \emph {et~al.}(2006)\citenamefont {Ronen},
  \citenamefont {Bortolotti},\ and\ \citenamefont {Bohn}}]{Ronen2006a}%
  \BibitemOpen
  \bibfield  {author} {\bibinfo {author} {\bibfnamefont {Shai}\ \bibnamefont
  {Ronen}}, \bibinfo {author} {\bibfnamefont {Daniele C.~E.}\ \bibnamefont
  {Bortolotti}}, \ and\ \bibinfo {author} {\bibfnamefont {John~L.}\
  \bibnamefont {Bohn}},\ }\bibfield  {title} {\enquote {\bibinfo {title}
  {{B}ogoliubov modes of a dipolar condensate in a cylindrical trap},}\ }\href
  {\doibase 10.1103/PhysRevA.74.013623} {\bibfield  {journal} {\bibinfo
  {journal} {Phys. Rev. A}\ }\textbf {\bibinfo {volume} {74}},\ \bibinfo
  {pages} {013623} (\bibinfo {year} {2006})}\BibitemShut {NoStop}%
\bibitem [{\citenamefont {Giovanazzi}\ \emph {et~al.}(2002)\citenamefont
  {Giovanazzi}, \citenamefont {G\"orlitz},\ and\ \citenamefont
  {Pfau}}]{Giovanazzi2002b}%
  \BibitemOpen
  \bibfield  {author} {\bibinfo {author} {\bibfnamefont {Stefano}\ \bibnamefont
  {Giovanazzi}}, \bibinfo {author} {\bibfnamefont {Axel}\ \bibnamefont
  {G\"orlitz}}, \ and\ \bibinfo {author} {\bibfnamefont {Tilman}\ \bibnamefont
  {Pfau}},\ }\bibfield  {title} {\enquote {\bibinfo {title} {Tuning the dipolar
  interaction in quantum gases},}\ }\href {\doibase
  10.1103/PhysRevLett.89.130401} {\bibfield  {journal} {\bibinfo  {journal}
  {Phys. Rev. Lett.}\ }\textbf {\bibinfo {volume} {89}},\ \bibinfo {pages}
  {130401} (\bibinfo {year} {2002})}\BibitemShut {NoStop}%
\bibitem [{\citenamefont {Tang}\ \emph {et~al.}(2018)\citenamefont {Tang},
  \citenamefont {Kao}, \citenamefont {Li},\ and\ \citenamefont
  {Lev}}]{Tang2018a}%
  \BibitemOpen
  \bibfield  {author} {\bibinfo {author} {\bibfnamefont {Yijun}\ \bibnamefont
  {Tang}}, \bibinfo {author} {\bibfnamefont {Wil}\ \bibnamefont {Kao}},
  \bibinfo {author} {\bibfnamefont {Kuan-Yu}\ \bibnamefont {Li}}, \ and\
  \bibinfo {author} {\bibfnamefont {Benjamin~L.}\ \bibnamefont {Lev}},\
  }\bibfield  {title} {\enquote {\bibinfo {title} {Tuning the dipole-dipole
  interaction in a quantum gas with a rotating magnetic field},}\ }\href
  {\doibase 10.1103/PhysRevLett.120.230401} {\bibfield  {journal} {\bibinfo
  {journal} {Phys. Rev. Lett.}\ }\textbf {\bibinfo {volume} {120}},\ \bibinfo
  {pages} {230401} (\bibinfo {year} {2018})}\BibitemShut {NoStop}%
\bibitem [{\citenamefont {Lu}\ \emph {et~al.}(2010)\citenamefont {Lu},
  \citenamefont {Lu}, \citenamefont {Zhang}, \citenamefont {Qiu}, \citenamefont
  {Pu},\ and\ \citenamefont {Yi}}]{Lu2010a}%
  \BibitemOpen
  \bibfield  {author} {\bibinfo {author} {\bibfnamefont {H.-Y.}\ \bibnamefont
  {Lu}}, \bibinfo {author} {\bibfnamefont {H.}~\bibnamefont {Lu}}, \bibinfo
  {author} {\bibfnamefont {J.-N.}\ \bibnamefont {Zhang}}, \bibinfo {author}
  {\bibfnamefont {R.-Z.}\ \bibnamefont {Qiu}}, \bibinfo {author} {\bibfnamefont
  {H.}~\bibnamefont {Pu}}, \ and\ \bibinfo {author} {\bibfnamefont
  {S.}~\bibnamefont {Yi}},\ }\bibfield  {title} {\enquote {\bibinfo {title}
  {Spatial density oscillations in trapped dipolar condensates},}\ }\href
  {\doibase 10.1103/PhysRevA.82.023622} {\bibfield  {journal} {\bibinfo
  {journal} {Phys. Rev. A}\ }\textbf {\bibinfo {volume} {82}},\ \bibinfo
  {pages} {023622} (\bibinfo {year} {2010})}\BibitemShut {NoStop}%
\bibitem [{\citenamefont {Bao}\ \emph {et~al.}(2010)\citenamefont {Bao},
  \citenamefont {Cai},\ and\ \citenamefont {Wang}}]{Bao2010a}%
  \BibitemOpen
  \bibfield  {author} {\bibinfo {author} {\bibfnamefont {Weizhu}\ \bibnamefont
  {Bao}}, \bibinfo {author} {\bibfnamefont {Yongyong}\ \bibnamefont {Cai}}, \
  and\ \bibinfo {author} {\bibfnamefont {Hanquan}\ \bibnamefont {Wang}},\
  }\bibfield  {title} {\enquote {\bibinfo {title} {Efficient numerical methods
  for computing ground states and dynamics of dipolar {B}ose-{E}instein
  condensates},}\ }\href {\doibase 10.1016/j.jcp.2010.07.001} {\bibfield
  {journal} {\bibinfo  {journal} {J. Comput. Phys.}\ }\textbf {\bibinfo
  {volume} {229}},\ \bibinfo {pages} {7874} (\bibinfo {year}
  {2010})}\BibitemShut {NoStop}%
\bibitem [{\citenamefont {Antoine}\ \emph {et~al.}(2017)\citenamefont
  {Antoine}, \citenamefont {Levitt},\ and\ \citenamefont
  {Tang}}]{Antoine2017a}%
  \BibitemOpen
  \bibfield  {author} {\bibinfo {author} {\bibfnamefont {Xavier}\ \bibnamefont
  {Antoine}}, \bibinfo {author} {\bibfnamefont {Antoine}\ \bibnamefont
  {Levitt}}, \ and\ \bibinfo {author} {\bibfnamefont {Qinglin}\ \bibnamefont
  {Tang}},\ }\bibfield  {title} {\enquote {\bibinfo {title} {Efficient spectral
  computation of the stationary states of rotating {B}ose-{E}instein
  condensates by preconditioned nonlinear conjugate gradient methods},}\ }\href
  {\doibase 10.1016/j.jcp.2017.04.040} {\bibfield  {journal} {\bibinfo
  {journal} {J. Comput. Phys.}\ }\textbf {\bibinfo {volume} {343}},\ \bibinfo
  {pages} {92} (\bibinfo {year} {2017})}\BibitemShut {NoStop}%
\bibitem [{\citenamefont {Knight}\ \emph {et~al.}(2019)\citenamefont {Knight},
  \citenamefont {Bland}, \citenamefont {Parker},\ and\ \citenamefont
  {Martin}}]{Knight2019a}%
  \BibitemOpen
  \bibfield  {author} {\bibinfo {author} {\bibfnamefont {Mitchell~J.}\
  \bibnamefont {Knight}}, \bibinfo {author} {\bibfnamefont {Thomas}\
  \bibnamefont {Bland}}, \bibinfo {author} {\bibfnamefont {Nick~G.}\
  \bibnamefont {Parker}}, \ and\ \bibinfo {author} {\bibfnamefont {Andy~M.}\
  \bibnamefont {Martin}},\ }\href@noop {} {\enquote {\bibinfo {title} {Improved
  low-dimensional wave equations for cigar-shaped and disk-shaped dipolar
  {B}ose-{E}instein condensates},}\ } (\bibinfo {year} {2019}),\ \Eprint
  {http://arxiv.org/abs/1908.02395} {arXiv:1908.02395 [cond-mat.quant-gas]}
  \BibitemShut {NoStop}%
\bibitem [{\citenamefont {Chomaz}\ \emph {et~al.}(2019)\citenamefont {Chomaz},
  \citenamefont {Petter}, \citenamefont {Ilzh\"ofer}, \citenamefont {Natale},
  \citenamefont {Trautmann}, \citenamefont {Politi}, \citenamefont
  {Durastante}, \citenamefont {van Bijnen}, \citenamefont {Patscheider},
  \citenamefont {Sohmen}, \citenamefont {Mark},\ and\ \citenamefont
  {Ferlaino}}]{Chomaz2019a}%
  \BibitemOpen
  \bibfield  {author} {\bibinfo {author} {\bibfnamefont {L.}~\bibnamefont
  {Chomaz}}, \bibinfo {author} {\bibfnamefont {D.}~\bibnamefont {Petter}},
  \bibinfo {author} {\bibfnamefont {P.}~\bibnamefont {Ilzh\"ofer}}, \bibinfo
  {author} {\bibfnamefont {G.}~\bibnamefont {Natale}}, \bibinfo {author}
  {\bibfnamefont {A.}~\bibnamefont {Trautmann}}, \bibinfo {author}
  {\bibfnamefont {C.}~\bibnamefont {Politi}}, \bibinfo {author} {\bibfnamefont
  {G.}~\bibnamefont {Durastante}}, \bibinfo {author} {\bibfnamefont {R.~M.~W.}\
  \bibnamefont {van Bijnen}}, \bibinfo {author} {\bibfnamefont
  {A.}~\bibnamefont {Patscheider}}, \bibinfo {author} {\bibfnamefont
  {M.}~\bibnamefont {Sohmen}}, \bibinfo {author} {\bibfnamefont {M.~J.}\
  \bibnamefont {Mark}}, \ and\ \bibinfo {author} {\bibfnamefont
  {F.}~\bibnamefont {Ferlaino}},\ }\bibfield  {title} {\enquote {\bibinfo
  {title} {Long-lived and transient supersolid behaviors in dipolar quantum
  gases},}\ }\href {\doibase 10.1103/PhysRevX.9.021012} {\bibfield  {journal}
  {\bibinfo  {journal} {Phys. Rev. X}\ }\textbf {\bibinfo {volume} {9}},\
  \bibinfo {pages} {021012} (\bibinfo {year} {2019})}\BibitemShut {NoStop}%
\bibitem [{\citenamefont {Tanzi}\ \emph
  {et~al.}(2019{\natexlab{b}})\citenamefont {Tanzi}, \citenamefont {Roccuzzo},
  \citenamefont {Lucioni}, \citenamefont {Fam{\`a}}, \citenamefont {Fioretti},
  \citenamefont {Gabbanini}, \citenamefont {Modugno}, \citenamefont {Recati},\
  and\ \citenamefont {Stringari}}]{Tanzi2019b}%
  \BibitemOpen
  \bibfield  {author} {\bibinfo {author} {\bibfnamefont {L.}~\bibnamefont
  {Tanzi}}, \bibinfo {author} {\bibfnamefont {S.~M.}\ \bibnamefont {Roccuzzo}},
  \bibinfo {author} {\bibfnamefont {E.}~\bibnamefont {Lucioni}}, \bibinfo
  {author} {\bibfnamefont {F.}~\bibnamefont {Fam{\`a}}}, \bibinfo {author}
  {\bibfnamefont {A.}~\bibnamefont {Fioretti}}, \bibinfo {author}
  {\bibfnamefont {C.}~\bibnamefont {Gabbanini}}, \bibinfo {author}
  {\bibfnamefont {G.}~\bibnamefont {Modugno}}, \bibinfo {author} {\bibfnamefont
  {A.}~\bibnamefont {Recati}}, \ and\ \bibinfo {author} {\bibfnamefont
  {S.}~\bibnamefont {Stringari}},\ }\bibfield  {title} {\enquote {\bibinfo
  {title} {Supersolid symmetry breaking from compressional oscillations in a
  dipolar quantum gas},}\ }\href {https://doi.org/10.1038/s41586-019-1568-6}
  {\bibfield  {journal} {\bibinfo  {journal} {Nature}\ }\textbf {\bibinfo
  {volume} {574}},\ \bibinfo {pages} {382} (\bibinfo {year}
  {2019}{\natexlab{b}})}\BibitemShut {NoStop}%
\bibitem [{\citenamefont {Guo}\ \emph {et~al.}(2019)\citenamefont {Guo},
  \citenamefont {B{\"o}ttcher}, \citenamefont {Hertkorn}, \citenamefont
  {Schmidt}, \citenamefont {Wenzel}, \citenamefont {B{\"u}chler}, \citenamefont
  {Langen},\ and\ \citenamefont {Pfau}}]{Guo2019a}%
  \BibitemOpen
  \bibfield  {author} {\bibinfo {author} {\bibfnamefont {Mingyang}\
  \bibnamefont {Guo}}, \bibinfo {author} {\bibfnamefont {Fabian}\ \bibnamefont
  {B{\"o}ttcher}}, \bibinfo {author} {\bibfnamefont {Jens}\ \bibnamefont
  {Hertkorn}}, \bibinfo {author} {\bibfnamefont {Jan-Niklas}\ \bibnamefont
  {Schmidt}}, \bibinfo {author} {\bibfnamefont {Matthias}\ \bibnamefont
  {Wenzel}}, \bibinfo {author} {\bibfnamefont {Hans~Peter}\ \bibnamefont
  {B{\"u}chler}}, \bibinfo {author} {\bibfnamefont {Tim}\ \bibnamefont
  {Langen}}, \ and\ \bibinfo {author} {\bibfnamefont {Tilman}\ \bibnamefont
  {Pfau}},\ }\bibfield  {title} {\enquote {\bibinfo {title} {The low-energy
  goldstone mode in a trapped dipolar supersolid},}\ }\href
  {https://doi.org/10.1038/s41586-019-1569-5} {\bibfield  {journal} {\bibinfo
  {journal} {Nature}\ }\textbf {\bibinfo {volume} {574}},\ \bibinfo {pages}
  {386} (\bibinfo {year} {2019})}\BibitemShut {NoStop}%
\bibitem [{\citenamefont {Natale}\ \emph {et~al.}(2019)\citenamefont {Natale},
  \citenamefont {van Bijnen}, \citenamefont {Patscheider}, \citenamefont
  {Petter}, \citenamefont {Mark}, \citenamefont {Chomaz},\ and\ \citenamefont
  {Ferlaino}}]{Natale2019a}%
  \BibitemOpen
  \bibfield  {author} {\bibinfo {author} {\bibfnamefont {G.}~\bibnamefont
  {Natale}}, \bibinfo {author} {\bibfnamefont {R.~M.~W.}\ \bibnamefont {van
  Bijnen}}, \bibinfo {author} {\bibfnamefont {A.}~\bibnamefont {Patscheider}},
  \bibinfo {author} {\bibfnamefont {D.}~\bibnamefont {Petter}}, \bibinfo
  {author} {\bibfnamefont {M.~J.}\ \bibnamefont {Mark}}, \bibinfo {author}
  {\bibfnamefont {L.}~\bibnamefont {Chomaz}}, \ and\ \bibinfo {author}
  {\bibfnamefont {F.}~\bibnamefont {Ferlaino}},\ }\bibfield  {title} {\enquote
  {\bibinfo {title} {Excitation spectrum of a trapped dipolar supersolid and
  its experimental evidence},}\ }\href {\doibase
  10.1103/PhysRevLett.123.050402} {\bibfield  {journal} {\bibinfo  {journal}
  {Phys. Rev. Lett.}\ }\textbf {\bibinfo {volume} {123}},\ \bibinfo {pages}
  {050402} (\bibinfo {year} {2019})}\BibitemShut {NoStop}%
\end{thebibliography}
%

\end{document}